\documentclass[10pt]{article}

\usepackage{graphicx}
\usepackage{graphics}
\usepackage{amscd}
\usepackage{amsmath}
\usepackage{amsfonts}
\usepackage{amssymb}
\usepackage{color}
\usepackage{cite}

\newcommand{\bF}{ {\mathbb F}}

\newcommand{\C}{ {\mathcal C}}

\newcommand{\red}{\color{red}}
\newtheorem{theorem}{Theorem}

\newtheorem{example}[theorem]{Example}

\newtheorem{lemma}[theorem]{Lemma}

\newtheorem{proposition}[theorem]{Proposition}
\newtheorem{remark}[theorem]{Remark}

\begin{document}

\title{A class of two or three weights linear codes \\ and their complete weight enumerators \footnote{
 *Corresponding author.\,\, E-Mail addresses:
 dzheng@hubu.edu.cn(D. Zheng),
zhaoqing9@126.com(Q. Zhao), waxiqq@163.com (X. Wang), Zhangyan@hubu.edu.cn(Y. Zhang)
}}

\author{ Dabin Zheng*$^1$, Qing Zhao$^1$, Xiaoqiang Wang$^1$ and Yan Zhang$^2$}

\date{\small 1. Hubei Province Key Laboratory of Applied Mathematics, \\
Faculty of Mathematics and Statistics, Hubei University, Wuhan 430062, China \\
2. School of Computer Science and Information Engineering, Hubei University, Wuhan 430062, China
}
\maketitle

\leftskip 1.0in
\rightskip 1.0in
\noindent {\bf Abstract.}
In the past few years, linear codes with few weights and their weight analysis have been widely studied.
In this paper, we further investigate a class of two-weight or three-weight linear codes from defining sets and determine their weight and complete weight enumerators
by application of the theory of quadratic forms and some special Weil sums over finite fields. Some punctured codes of the discussed linear codes are optimal or almost optimal
with respect to the Griesmer bound. This paper generalizes some results in~\cite{ZhuXu2017,Jian2019}.


\vskip 6pt
\noindent {\it Keywords:} Linear code, Weight enumerator, Complete weight enumerator, Exponential sum, Quadratic form
\vskip 6pt
\noindent {\it 2010 Mathematics Subject Classification:} 94B05, 94B15

\vskip 35pt

\leftskip 0.0in
\rightskip 0.0in

\section{Introduction}

Let $\bF_p$ be a finite field of size $p$, where $p$ is an odd prime.
An $[n, k, d]$ linear code $\C$ over $\bF_p$ is a $k$-dimensional
subspace of $\bF_p^n$ with minimum distance $d$. Let $A_i$ denote the number of codewords
with Hamming weight $i$ in $\C$. The weight enumerator of $\C$ is defined by
$1+ A_1 x+ A_2x^2 + \cdots + A_nx^n$ and the sequence $(1, A_1, A_2, \cdots, A_n)$ is called the weight distribution of $\C$. If the number of nonzero $A_i$ in this sequence is equal to $t$, then we call $\C$ a $t$-weight code.
The weight distribution of a code not only gives the error correcting ability of the code, but also allows the computation of the error probability of error detection and correction~\cite{Klove2007}. So, the study of the weight distribution of a linear code is important in both theory and applications. Linear codes with few weights attracts many researchers' attention due to their wide applications in secret schemes, strongly regular graphs,
association schemes and authentication codes. The recent progress on constructions of two-weight and three-weight linear codes can be seen in~\cite{DDing2015, DingLietal2015, HengYue2016, HengYueLi2016, HengYue2017, Jian2019, Lietal2016, LuoCaoetal2018, Tan2018, Tang2018, TangLietal2016, Wangetal2016, Xiaetal2017, ZhouDing2013, ZhouDing2014, ZhouLietal2015} and the references therein.

The complete weight enumerator of a code $\mathcal{C}$ over $\mathbb{F}_{p}$ enumerates the codewords according to the number of symbols of each kind contained in each codeword. Denote elements in the field by
$\bF_{p}=\{0, 1,\cdots, p-1\}$. The composition of a vector ${\bf v} = (v_0, v_1, \cdots, v_{n-1})\in \bF_p^n$ is defined to be
comp(${\bf v}$) $= (k_0, k_1, \cdots, k_{p-1})$, where each $k_i =k_i({\bf v})$ is the number of components $v_j (0\leq j \leq n-1)$ of ${\bf v}$ that are equal to $i$. It is clear that $\sum_{i=0}^{p-1} k_i =n$. For a codeword $\mathbf{c}=(c_0, c_1, \cdots, c_{n-1})\in\mathcal{C}$, the complete weight enumerator of $\mathbf{c}$ is the monomial
\[ w(\mathbf{c})= w_0^{k_0(\mathbf{c})}  w_1^{k_1(\mathbf{c})} \cdots w_{p-1}^{k_{p-1}(\mathbf{c})}\]
in the variables $w_0, w_1, \cdots, w_{p-1}$, where $k_i(\mathbf{c}) (0 \leq i\leq  p-1)$ denotes the number of components $c_j$ of that equals~$i$. The complete weight enumerator of a linear code  $\mathcal{C}$ is the homogeneous polynomial,
\[CWE(\mathcal{C})=\sum_{\mathbf{c}\in\mathcal{C}} w_0^{k_0(\mathbf{c})}  w_1^{k_1(\mathbf{c})} \cdots w_{p-1}^{k_{p-1}(\mathbf{c})}
                =\sum_{(k_0, k_1, \cdots, k_{p-1})\in B_n}A(k_0, k_1, \cdots, k_{p-1})  w_0^{k_0}  w_1^{k_1} \cdots w_{p-1}^{k_{p-1}},
\]
where $B_n=\left\{ (k_0, k_1, \cdots, k_{p-1}) \, :\, 0\leq k_i\leq n, \, \sum_{i=0}^{p-1} k_i =n \right\}$ and $A(k_0, k_1, \cdots, k_{p-1})$ denotes the number of codewords
$\mathbf{c}\in \C$ with comp($\mathbf{c}$) $= (k_0, k_1, \cdots, k_{p-1})$.

The complete weight enumerators of linear codes not only give the weight enumerators but also demonstrate the frequency of each symbol appearing in each codeword. They have wide applications such as
in authentication codes~\cite{DingWang2005,DingHellesethetal2007}, constant composition codes~\cite{DingC2008,DingYin2005} and the computation of Walsh transform values~\cite{HellesethKholosha2006}.
So, it is interesting to determine complete weight enumerators of linear codes. The complete weight enumerators of Reed-Solomon codes were obtained by Blake and Kith~\cite{Blake1991} and
Kuzmin and Nechaev~\cite{Kuzmin2001} studied the complete weight enumerators of the generalized Kerdock code and related linear codes over Galois rings. Recently, there are many works on complete
weight enumerators of specific linear codes. The reader is referred to~\cite{AhnKaLi2017, LiYueFu2016, LuoG2018, WuY2019, YangShu2019, YangKong2017, LiuYi2018, XuGuang2018, LiBae2016} and the references therein.

Let $\bF_{p^m}$ be a finite field with $p^m$ elements, where $m$ is a positive integer. Let ${\rm Tr}$ denote the trace function from $\bF_{p^m}$ to $\bF_p$.
From a subset $D=\{d_1, d_2, \ldots, d_n\}\subset \bF_{p^m}$, Ding et al.~\cite{DingNieder2007} first defined a generic class of linear codes of length $n=|D|$ over $\bF_p$ as
\begin{equation*}
\C_D = \left\{ \left( {\rm Tr}( xd_1), {\rm Tr}( xd_2), \cdots, {\rm Tr}(xd_n) \right) \, |\, x\in \bF_{p^m} \right\}.
\end{equation*}
Here, $D$ is called the defining set of $\C_D$. This construction is generic in the sense that many classes of known codes could be produced by selecting the defining set $D$.
By application of this technique, many good linear codes with few weights have been constructed~\cite{Ding2015,Ding2016, DingNieder2007, DDing2015, DingLietal2015, HengYue2016, HengYueLi2016, HengYue2017, HengYue20162, Lietal2016, LuoCaoetal2018, Xiaetal2017, ZhengBao2017}.
Motivated by this construction, Zhu et al.~\cite{ZhuXu2017} studied the weight distribution of the linear code
\begin{equation}\label{defcode}
\begin{split}
\mathcal{C}_{D}=\left\{c(a,b)=\left({\rm Tr}(ax+by)\right)_{(x ,y)\in {D}}: \, a,b \in \mathbb{F}_{p^m} \right\},
\end{split}
\end{equation}
where
\begin{equation}\label{defset}
D=\left\{(x,y)\in \mathbb{F}_{p^m}^2\setminus\{(0,0)\}:\,\, {\rm Tr}(x^{{p^k}+1}+y^{{p^\ell}+1})=c,c\in \mathbb{F}_p\right\}
\end{equation}
for $m/(m,k)$ being odd and $m/(m,\ell)$ being even. Almost at the same time, Jian et al.~\cite{Jian2019} also considered the linear code $\C_D$ in (\ref{defcode}) for the case $k=0$ and $c=0$.

In this paper, we first study the weight distribution of the linear code $\C_D$ for any positive integers $m, k$ and $\ell$ by careful analysis of ranks of the discussed quadratic form, and generalize the results in~\cite{ZhuXu2017,Jian2019}.
Secondly, we determine the complete weight enumerators of the linear code $\C_D$ by application of quadratic form theory over finite fields. Moreover, the punctured code of $\C_D$ is discussed and some optimal
or almost optimal linear codes with respect to the Griesmer bound are obtained.


The rest of this paper is organized as follows. In  Section~2, we introduce some preliminaries, which will be used in the following sections. Section~3 investigates the weight and complete weight enumerators of the linear code $\C_D$. In Section~4, the punctured version of $\C_D$ is discussed. Section~5 concludes this paper.

\section{Preliminaries}

Throughout this paper, we adopt the following notation unless otherwise stated:
\begin{description}
\item{$\bullet$} $\bF_{p^m}$ is a finite field with $p^m$ elements.
\item{$\bullet$} Tr($\cdot$) is the trace function from $\bF_{p^m}$ to $\bF_p$.
\item{$\bullet$} $v_2(\cdot)$ is the 2-adic order function and we denote $v_2(0)=\infty$.
\item{$\bullet$} $k$ and $\ell$ are positive integers, $\gcd(k,m)=u$ and  $\gcd(\ell,m)=v$.
\item{$\bullet$} $\zeta_p=e^{\frac{2\pi\sqrt{-1}}{p}}$ is the primitive $p$-th root of unity.
\end{description}

Let $\psi$ be a multiplicative character of $\mathbb{F}_{p^m}^*$. The Gaussian sum $G(\psi)$ is defined by
\[ G(\psi) = \sum_{x\in \mathbb{F}_{p^m}^*} \psi(x)\chi(x),\]
where $\chi$ be the canonical additive character of $\mathbb{F}_{p^m}$. The explicit values of Gaussian sums are very difficult to determine and are known for only a few cases.
Let $\eta_m$ be the quadratic multiplicative character of $\mathbb{F}_{p^m}^*$ and $G_m$ denote the Gaussian sum $G(\eta_m)$ for short. Particularly, $G_1$ is the Gaussian
sum $G(\eta)$ over $\bF_p^*$, where $\eta$ is the quadratic multiplicative character of $\mathbb{F}_{p}^*$.
\begin{lemma}\cite[Theorem 5.15]{Lidl1983}\label{lem1}
Let $\mathbb{F}_{p^m}$ be a finite field with $p^m$ elements and $\eta_m$ be the quadratic multiplicative character of $\mathbb{F}_{p^m}^*$. Then
\[G_m= (-1)^{m-1} \sqrt{(-1)^{\frac{(p-1)m}{2}}p^m}=\left\{ \begin{array}{lcl}
           (-1)^{m-1}p^{\frac{m}{2}}, & {\rm if}\, \,\, p\equiv 1 \pmod 4, \vspace{4mm}\\
           (-1)^{m-1}(\sqrt{-1})^mp^{\frac{m}{2}}, & {\rm if}\, \,\, p\equiv 3 \pmod 4. \end{array}  \right.\]
\end{lemma}
In particular, $G_1=(-1)^{\frac{p-1}{4}}p^{\frac{1}{2}}$.

By identifying the finite field $\mathbb{F}_{p^m}$ with an $m$-dimensional vector space
$\mathbb{F}_{p}^m$ over $\mathbb{F}_{p}$, a function~$f$ from $\mathbb{F}_{p^m}$ to $\mathbb{F}_p$ can be viewed as an $m$-variable polynomial over $\mathbb{F}_{p}$.
The function $f(x)$ is called a quadratic form
if it is a homogenous polynomial of degree two as follows:
\[f(x_1, x_2, \cdots, x_m) = \sum_{1\leq i\leq j\leq m} a_{ij} x_ix_j , \,\, \, a_{ij}\in \mathbb{F}_{p},\]
where we fix  a basis of $\mathbb{F}_p^m$ over $\mathbb{F}_p$ and identify $x\in \mathbb{F}_{p^m}$ with a vector $(x_1, x_2, \cdots, x_m)\in \mathbb{F}_p^m$.
The rank of the quadratic form $f(x)$ is defined as the codimension of the $\mathbb{F}_p$-vector space
\[ V = \left\{ x\in \mathbb{F}_p^m\, \, |\,\, f(x+z)-f(x)-f(z) = 0, \,\, \mbox{for all}\,\, z\in \mathbb{F}_{p}^m \right\} ,\]
which is denoted by rank$(f)$. Then $|V| = p^{m-{\rm rank}(f)}$.

\vskip 6pt
For a quadratic form $Q(x)$ with $m$ variables over $\mathbb{F}_p$, there exists a symmetric matrix $A$ such that $Q(x)=XAX^\prime$, where $X=(x_1, x_2, \cdots, x_m)\in \mathbb{F}_p^m$ and $X^\prime$ denotes the transpose of $X$. The determinant ${\rm det}(Q)$ of $Q(x)$ is defined to be the determinant of $A$, and $Q(x)$ is nondegenerate if ${\rm det}(Q)\neq 0$. It is known that there exists a nonsingular matrix $T$ such that $TAT^\prime$ is a diagonal matrix \cite{Lidl1983}. Making a nonsingular linear substitution $X=YT$ with $Y=(y_1, y_2, \cdots, y_m)$, we have
\[ Q(x)= YTAT^\prime Y^\prime = \sum_{i=1}^r a_i y_i^2, \,\, a_i\in \mathbb{F}_p ,\]
where $r(\leq m)$ is the rank of $Q(x)$. The following lemma gives a general result on the exponential sums of a quadratic function from $\mathbb{F}_{p^m}$ to $\mathbb{F}_p$.
These sums are also known as Weil sums.

\begin{lemma}\cite[Theorems 5.15 and 5.33]{Lidl1983}\label{lem:quadraticsum}
Let $Q(x)$ be a quadratic function from $\mathbb{F}_{p^m}$ to $\mathbb{F}_p$ with rank $r(r\neq0)$, and
$\eta$ be the quadratic multiplicative character of $\mathbb{F}_p^*$. Then
\[  \sum_{x\in \mathbb{F}_{p^m}} \zeta_p^{Q(x)}=\left\{ \begin{array}{lcl}
          \eta(\Delta)p^{m-\frac{r}{2}}, & {\rm if} \,\, p \equiv 1 \,\, ({\rm mod} \,\, 4), \\ \\
           (-1)^{\frac{r}{2}}\eta(\Delta)p^{m-\frac{r}{2}}, & { \rm if} \,\, p \equiv 3 \,\, ({\rm mod} \,\, 4), \end{array}  \right.\]
where $\Delta$ is the determinant of $Q(x)$. Furthermore, for any $z\in \mathbb{F}_{p}^*$,
\begin{equation*}
\sum_{x\in \mathbb{F}_{p^m}} \zeta_p^{z Q(x)}=\eta_m^r(z)\sum_{x\in \mathbb{F}_{p^m}} \zeta_p^{ Q(x)},
\end{equation*}
\end{lemma}
where $\eta_m$ is the quadratic multiplicative character of $\mathbb{F}_{p^m}^*$.

\begin{lemma}\cite[Theorems 5.33]{Lidl1983}\label{lem:functions}
 Let $f(x)=a_2x^2+a_1x+a_0\in\mathbb{F}_{p^m}[x]$ with $a_2\ne0$. Then
\begin{equation*}
\sum_{x\in \mathbb{F}_{p^m}} \zeta_p^{{\rm Tr}\left(f(x)\right)}=\zeta_p^{{\rm Tr}\left(a_0-a_1^2(4a_2)^{-1}\right)}\eta_m(a_2)G_m,
\end{equation*}
where $G_m$ is the Gaussian sum $G(\eta_m)$.
\end{lemma}

\begin{lemma}\cite{Draper2007}\label{lem:rank}
Let ${\rm Tr}(x^{p^k+1})$ be the quadratic function from $\mathbb{F}_{p^m}$ to $\mathbb{F}_p$ with rank $r$. Then
\[r =\left\{ \begin{array}{lll}
           m-2u, & {\rm if}\, \,\, v_2(m)> v_2(k) +1, \vspace{3mm} \\
           m, & \, \, otherwises, \end{array}  \right.\]
where $u=\gcd(m,k)$.
\end{lemma}

The weight and complete weight enumerator of the linear code $\C_D$ is related to the rank of the quadratic form ${\rm Tr}(x^{p^k+1}+y^{p^\ell +1})$ and
the following Weil sum
\begin{equation}\label{eq:sab}
S_k(a,b)=\sum_{x \in \mathbb{F}_{p^m}}\chi\left(ax^{p^k+1}+bx\right), a\in \mathbb{F}_{p^m}^*,b\in \mathbb{F}_{p^m}.
\end{equation}
When $b=0$, the Weil sum $S_k(a, 0)$ is as follows.
\begin{lemma}\cite[Corollary 7.6]{Draper2007}\label{lem:ska0}
Let $v_2(\cdot)$ denote the 2-adic order function and $v_2(0)=\infty$. Let $\eta_m$ be the quadratic multiplicative character of $\mathbb{F}_{p^m}^*$. For $ a\in \mathbb{F}_{p^m}^* $, the following
results hold.
\begin{description}
\item{{\rm (i)}} If $v_2(m)\leq v_2(k)$, then
\[ S_k(a,0)=\eta_m(a)(-1)^{m-1}(\sqrt{-1})^{\frac{(p-1)^2m}{4}}p^{\frac{m}{2}}.\]
\item{{\rm (ii)}} If $v_2(m)=v_2(k)+1$, then
\[S_k(a,0)=\left\{ \begin{array}{lll}
            p^{\frac{m+\gcd(2k,m)}{2}},& {\rm if}\,\,a^{\frac{(p^k-1)(p^m-1)}{p^{\gcd(2k,m)}-1}}=-1, \vspace{3mm}\\
            -p^{\frac{m}{2}}, & \,\, otherwise. \end{array}  \right.\]
\item{{\rm (iii)}} If $v_2(m)>v_2(k)+1$, then
\[ S_k(a,0)=\left\{ \begin{array}{lll}
    -p^{\frac{m+\gcd(2k,m)}{2}},& {\rm if}\,\,a^{\frac{(p^k-1)(p^m-1)}{p^{\gcd(2k,m)}-1}}=1, \vspace{3mm} \\
     p^{\frac{m}{2}} , & \,\, otherwise. \end{array}  \right.\]
\end{description}
\end{lemma}

When $b\neq 0$, the value of $S_k(a, b)$ is related to the solutions of  the polynomial $a^{p^k}x^{p^{2k}}+ax=0$.
We first recall the result on this equation.

\begin{lemma}\cite[Theorem 4.1]{Coulter19980}\label{lemGxn}
Let $ m, k$ be positive integers with $u = \gcd(m, k)$ and $a \in \bF_{p^m}^*$. The equation
$$a^{p^k}x^{p^{2k}}+ax=0$$
is solvable in $\mathbb{F}_{p^m}^*$ if and only if ${\frac{m}{u}}$ is even and $a^{\frac{p^m-1}{p^{u}+1}}=(-1)^{\frac{m}{2u}}$. In such cases, there are $p^{2u}-1$ non-zero solutions.
\end{lemma}

By application of this lemma, R. S.\ Coulter determined the possible values of the Weil sum $S_k(a, b)$ as follows.

\begin{lemma}\cite[Theorem 1]{Coulter1998}\label{lem:skab}
Let $m, k$ be positive integers with $u = \gcd(m, k)$. Let $a \in \mathbb{F}_{p^m}^*$  and $f(x)=a^{p^k}x^{p^{2k}}+ax$ be a permutation polynomial over $\mathbb{F}_{p^m}$. Assume that $x_0$ is the unique solution of the equation $f(x)=-b^{p^k}$. The following statements hold.
\begin{description}
\item{{\rm (i)}} If ${\frac{m}{u}}$ is odd, then
\begin{equation*}
\begin{split}
S_k(a,b)&=(-1)^{m-1}(\sqrt{-1})^{\frac{(p-1)^2}{4}3m} p^{\frac{m}{2}}\eta_m(a)\overline{\chi(ax_0^{p^k+1})} \vspace{3mm} \\
        &=\left\{ \begin{array}{lcl}
           (-1)^{m-1}p^{\frac{m}{2}}\eta_m(a)\overline{\chi(ax_0^{p^k+1})}, & {\rm if}\, \,\, p\equiv 1 \pmod 4, \vspace{3mm}\\
           (-1)^{m-1}(\sqrt{-1})^{3m}p^{\frac{m}{2}}\eta_m(a)\overline{\chi(ax_0^{p^k+1})}, & {\rm if}\, \,\, p\equiv 3 \pmod 4. \end{array}  \right.\\
\end{split}
\end{equation*}
\item{{\rm (ii)}} If ${\frac{m}{u}}$ is even, then $a^{\frac{p^m-1}{p^{u}+1}}\ne(-1)^{\frac{m}{2u}}$ and
 $$S_k(a,b)=(-1)^{\frac{m}{2u}}p^\frac{m}{2}\overline{\chi(ax_0^{p^k+1})},$$
\end{description}
where $\eta_m$ is the quadratic multiplicative of $\mathbb{F}_{p^m}^*$ and  $\chi$ is the canonical additive character of $\mathbb{F}_{p^m}$.
\end{lemma}

\begin{lemma}\cite[Theorem 2]{Coulter1998}\label{lem:skab2}
Let $m, k$ be positive integers with $u = \gcd(m, k)$ and $m$ being even. Assume that $f(x)=a^{p^k}x^{p^{2k}}+ax$ is not a permutation polynomial over $\mathbb{F}_{p^m}$, then ${S_k(a,b)=0}$ unless the equation $f(x)=-b^{p^k}$ is solvable. If the equation has a solution $x_0$, then
 $$ S_k(a,b)=-(-1)^{\frac{m}{2u}}p^{\frac{m}{2}+u}\overline{\chi(a x_0^{p^k+1})}.$$
\end{lemma}

For later use, we need the following lemma.

\begin{lemma}\cite[Lemma 13]{Jian2019}\label{lem:numsl}
Let $u=\gcd(m,k)$. If ${\frac{m}{u}} \equiv 0 \,\, {\rm mod}\,\, 4$, then
$$\left|\left\{c\in \mathbb{F}_{p^m}:\,\, x^{p^{2k}}+x=c^{p^{k}}\, \,{\rm is \,\,solvable \,\,in\,\,} \mathbb{F}_{p^m}\right\}\right|=p^{m-2u}.$$
\end{lemma}

In order to obtain the multiplicity of each weight of the discussed linear codes, we need the Pless power moment
identities on linear codes. Let $\mathcal{C}$ be an $[n, k]$ code over $\mathbb{F}_p$, and denote its dual by $\mathcal{C}^{\perp}$. Let
$A_i$ and $A^{\perp}_i$ be the number of codewords of weight $i$ in $\mathcal{C}$ and $\mathcal{C}^{\perp}$, respectively.
The first two Pless power moment identities are as follows (\cite{MacWilliam1997}, p. 131):
\begin{equation*}
\begin{split}
&\sum_{i=0}^nA_i=p^k;\\
&\sum_{i=0}^niA_i=p^{k-1}(pn-n-A_1^{\perp}).\\
\end{split}
\end{equation*}
For the linear code $\C_D$ defined in (\ref{defcode}) with the defining set $D$ in (\ref{defset}), it is easy to verify that $A_1^{\perp}=0$ if $(0,0) \not\in D$ from the
non-degenerate property of a trace function.

The following lemma on the bound of linear code is well-known.

\begin{lemma}(Griesmer Bound)\label{lem11}
If an $[n,k,d]$ $p$-ary code exists, then
\begin{equation*}
n\geq \sum_{i=0}^{k-1}\lceil\frac{d}{p^i}\rceil,
\end{equation*}
where the symbol $\lceil x \rceil$ denotes the smallest integer greater than or equal to $x$.
\end{lemma}

\section{Main results}

In this section, we investigate the weight enumerator and complete weight enumerator of the linear code $\mathcal{C}_D$ defined in (\ref{defcode}), where the defining set $D$ is given in (\ref{defset}). Firstly, we determine
the length of $\C_D$. Let $m, k$ and $\ell$ be integers with $u=\gcd(m,k)$ and $v=\gcd(m, \ell)$. For convenience, we define the following symbols.
\begin{equation}\label{equan03}
\begin{split}
\varepsilon_u=\left\{ \begin{array}{lcl}
           1, & {\rm if}\, \,\, v_2(m)>v_2(u)+1, \vspace{3mm}\\
           0, & {\rm if}\, \,\, v_2(m)\leq v_2(u)+1, \end{array}  \right.
\end{split} \quad
\begin{split}
\varepsilon_v=\left\{ \begin{array}{lcl}
           1, & {\rm if}\, \,\, v_2(m)>v_2(v)+1, \vspace{3mm} \\
           0, & {\rm if}\, \,\, v_2(m)\leq v_2(v)+1. \end{array}  \right.
\end{split}
\end{equation}

\begin{proposition}\label{pro:ww1} Let $\mathcal{C}_D$ be a linear code defined in (\ref{defcode}) with the defining set $D$ given in (\ref{defset}). Let $n=|D|$ be the length
of $\C_D$. If $c=0$, then
\begin{equation*}
\begin{split}
n=\begin{cases}
p^{2m-1}+(-1)^{\frac{(p-1)m}{2}}(p-1)p^{m-1}-1, & \text{if \,\,$2v_2(m)= v_2(u)+v_2(v)$,} \vspace{3mm} \\
p^{2m-1}+(-1)^{\frac{(p-1)m}{4}}(p-1)p^{m-1}-1, & \text{if \,\,$2v_2(m)=v_2(u)+v_2(v)+1$,} \vspace{3mm}\\
p^{2m-1}+p^{m+\varepsilon_uu+\varepsilon_vv}-p^{m+\varepsilon_uu+\varepsilon_vv-1}-1, & \text{if \,\,$2v_2(m)>v_2(u)+v_2(v)+1$.} \\
\end{cases}
\end{split}
\end{equation*}
If $c\in \mathbb{F}_p^*$, then
\begin{equation*}
\begin{split}
n=\begin{cases}
p^{2m-1}-(-1)^{\frac{(p-1)m}{2}}p^{m-1}, & \text{if \,\,$2v_2(m)= v_2(u)+v_2(v)$,} \vspace{3mm} \\
p^{2m-1}-(-1)^{\frac{(p-1)m}{4}}p^{m-1}, & \text{if \,\,$2v_2(m)=v_2(u)+v_2(v)+1$,}  \vspace{3mm}\\
p^{2m-1}-p^{m+\varepsilon_uu+\varepsilon_vv-1}, & \text{if \,\,$2v_2(m)>v_2(u)+v_2(v)+1$.} \\
\end{cases}
\end{split}
\end{equation*}
\end{proposition}
{\it Proof.} By the orthogonal property of additive characters and Lemma~\ref{lem:quadraticsum}, we have
\begin{equation*}
\begin{split}
 n&={\frac{1}{p}}\sum_{(x,y) \in \mathbb{F}_{p^m}^2\setminus\{(0,0)\}}\sum_{z \in \mathbb{F}_{p}}\zeta_p^{z\left({\rm Tr}(x^{p^k+1}+y^{p^\ell+1})-c\right)} \\
&={\frac{1}{p}}\sum_{x,y \in \mathbb{F}_{p^m}}\sum_{z \in \mathbb{F}_{p}}\zeta_p^{z\left({\rm Tr}(x^{p^k+1}+y^{p^\ell+1})-c\right)}-\frac{1}{p}\sum_{z \in \mathbb{F}_{p}}\zeta_p^{-zc} \\
&=p^{2m-1}+{\frac{1}{p}}\sum_{z \in \mathbb{F}_{p}^*}\sum_{x,y \in \mathbb{F}_{p^m}}\zeta_p^{z\left({\rm Tr}(x^{p^k+1}+y^{p^\ell+1})-c\right)}-\frac{1}{p}\sum_{z \in \mathbb{F}_{p}}\zeta_p^{-zc}  \\
&=p^{2m-1}+{\frac{1}{p}}\sum_{z \in \mathbb{F}_{p}^*}\zeta_p^{-zc}\sum_{x \in \mathbb{F}_{p^m}}\zeta_p^{z{\rm Tr}(x^{p^k+1})}\sum_{y \in \mathbb{F}_{p^m}}\zeta_p^{z{\rm Tr}(y^{p^\ell+1})}-\frac{1}{p}\sum_{z \in \mathbb{F}_{p}}\zeta_p^{-zc}  \\
&=p^{2m-1}+{\frac{1}{p}}\sum_{z \in \mathbb{F}_{p}^*}\zeta_p^{-zc}\sum_{x \in \mathbb{F}_{p^m}}\eta_m^{r_1}(z)\zeta_p^{{\rm Tr}(x^{p^k+1})}\sum_{y \in \mathbb{F}_{p^m}}\eta_m^{r_2}(z)\zeta_p^{{\rm Tr}(y^{p^\ell+1})}-\frac{1}{p}\sum_{z \in \mathbb{F}_{p}}\zeta_p^{-zc}\\
&=p^{2m-1}+{\frac{1}{p}}\sum_{z \in \mathbb{F}_{p}^*}\zeta_p^{-zc}\eta_m^{\left(r_1+r_2\right)}(z)\sum_{x \in \mathbb{F}_{p^m}}\zeta_p^{{\rm Tr}(x^{p^k+1})}\sum_{y \in \mathbb{F}_{p^m}}\zeta_p^{{\rm Tr}(y^{p^\ell+1})}-\frac{1}{p}\sum_{z \in \mathbb{F}_{p}}\zeta_p^{-zc}, \\
\end{split}
\end{equation*}
where $\eta_m$ is the quadratic multiplicative character of $\mathbb{F}_{p^m}^*$, and $r_1$ and $r_2$ are the ranks of the quadratic forms ${\rm Tr}(x^{p^k+1})$ and ${\rm Tr}(y^{p^\ell+1})$, respectively.
By Lemma~\ref{lem:rank}, it is easy to see that $\eta_m^{r_1+r_2}(z)=1$ for any $z \in \mathbb{F}_p^*$ since $r_1+r_2$ is even. Hence,
\begin{equation}\label{eq:6}
\begin{split}
n &=p^{2m-1}+{\frac{1}{p}}\sum_{z \in \mathbb{F}_{p}^*}\zeta_p^{-zc}\sum_{x \in \mathbb{F}_{p^m}}\zeta_p^{{\rm Tr}(x^{p^k+1})}\sum_{y \in \mathbb{F}_{p^m}}\zeta_p^{{\rm Tr}(y^{p^\ell+1})}-\frac{1}{p}\sum_{z \in \mathbb{F}_{p}}\zeta_p^{-zc} \vspace{4mm}\\
&= \left\{ \begin{array}{lll}
p^{2m-1} + \frac{p-1}{p} \Omega -1, &\quad {\rm if }\,\, c=0, \vspace{3mm} \\
p^{2m-1} -\frac{1}{p} \Omega, &\quad {\rm if } \,\, c\in \mathbb{F}_p^* ,\end{array} \right.
\end{split}
\end{equation}
where
\[\Omega=\sum_{x \in \mathbb{F}_{p^m}}\zeta_p^{{\rm Tr}(x^{p^k+1})}\sum_{y \in \mathbb{F}_{p^m}}\zeta_p^{{\rm Tr}(y^{p^\ell+1})}.
\]
The desired conclusion then follows from $(5)$ and Lemma \ref{lem:ska0}. $\square$

For any  $(a,b)\in \mathbb{F}_{p^m}^2\setminus\{(0,0)\}$, a nonzero codeword in $\C_{D}$ has the form $\mathbf{c}(a,b)=\left({\rm Tr}(ax+by)\right)_{(x,y)\in D}$, where $D$ is defined in (\ref{defset}).
Let $N(c,\rho)$ denote the number of components ${\rm Tr}(ax+by)$ of ${ \mathbf{c}(a,b)}$ which are equal to $\rho$, where $\rho\in\mathbb{F}_{p}$, i.e.,
\begin{equation*}
N(c,\rho)=\left|\left\{(x,y)\in \mathbb{F}_{p^m}^2\setminus\{(0,0)\}:\,\, {\rm Tr}\left(x^{{p^k}+1}+y^{{p^\ell}+1}\right)=c, {\rm Tr}\left(ax+by\right)=\rho\right\}\right|.
\end{equation*}
The Hamming weight of ${ \mathbf{c}({a,b})}$ is as follows:
\begin{equation}\label{eq:wt}
\begin{split}
{\rm wt_H}( \mathbf{c}(a,b))&=\sum_{\rho \in \mathbb{F}_{p^*}}N(c,\rho)=n-N(c,0),
\end{split}
\end{equation}
where $n$ is the length of linear code $\C_{D}$. Assume that $\eta_m$ is the quadratic multiplicative character of $\mathbb{F}_{p^m}^*$, and $r_1$ and $r_2$ are the ranks of the quadratic forms of ${\rm Tr}(y^{p^k+1})$ and
${\rm Tr}(y^{p^\ell+1})$, respectively. By Lemma~\ref{lem:quadraticsum},
\begin{equation}\label{eq:nab}
\begin{split}
N(c,\rho)&=\sum_{(x,y) \in \mathbb{F}_{p^m}^2\setminus \{(0,0)\}} \left({\frac{1}{p}}\sum_{z_1 \in \mathbb{F}_{p}}\zeta_p^{z_1({\rm Tr}(x^{p^k+1}+y^{p^\ell+1})-c)}\right)\left({\frac{1}{p}}\sum_{z_2 \in \mathbb{F}_{p}}\zeta_p^{z_2\left({\rm Tr}(ax+by)-\rho\right)}\right)\\
&=\sum_{x,y \in \mathbb{F}_{p^m}} \left({\frac{1}{p}}\sum_{z_1 \in \mathbb{F}_{p}}\zeta_p^{z_1({\rm Tr}(x^{p^k+1}+y^{p^\ell+1})-c)}\right)\left({\frac{1}{p}}\sum_{z_2 \in \mathbb{F}_{p}}\zeta_p^{z_2({\rm Tr}(ax+by)-\rho)}\right)-\frac{1}{p^2}\sum_{z_1,z_2 \in \mathbb{F}_{p}}\zeta_p^{-\left(z_1c+z_2\rho\right)}\\
&=p^{2m-2}+{\frac{1}{p^2}}\Bigg(\sum_{z_1 \in \mathbb{F}_{p^*}}\zeta_p^{-z_1c}\sum_{x \in \mathbb{F}_{p^m}}\zeta_p^{z_1{\rm Tr}(x^{p^k+1})}\sum_{y \in \mathbb{F}_{p^m}}\zeta_p^{z_1{\rm Tr}(y^{p^\ell+1})}+\sum_{z_2 \in \mathbb{F}_{p}^*}\zeta_p^{-z_2\rho}\sum_{x,y \in \mathbb{F}_{p^m}}\zeta_p^{z_2{\rm Tr}(ax+by)}\\
&+\sum_{z_1 \in \mathbb{F}_{p}^*}\zeta_p^{-z_1c}\sum_{z_2 \in \mathbb{F}_{p}^*}\zeta_p^{-z_2\rho}\sum_{x \in \mathbb{F}_{p^m}}\zeta_p^{{\rm Tr}(z_1x^{p^k+1}+z_2ax)}\sum_{y \in \mathbb{F}_{p^m}}\zeta_p^{{\rm Tr}(z_1y^{p^\ell+1}+z_2by)}\Bigg)-\frac{1}{p^2}\sum_{z_1,z_2 \in \mathbb{F}_{p}}\zeta_p^{-(z_1c+z_2\rho)}\\
&=p^{2m-2}+{\frac{1}{p^2}}\Bigg(\sum_{z_1 \in \mathbb{F}_{p^*}}\zeta_p^{-z_1c}\sum_{x \in \mathbb{F}_{p^m}}\eta_m^{r_1}(z_1)\zeta_p^{{\rm Tr}(x^{p^k+1})}\sum_{y \in \mathbb{F}_{p^m}}\eta_m^{r_2}(z_1)\zeta_p^{{\rm Tr}(y^{p^\ell+1})}\\
&+\sum_{z_1 \in \mathbb{F}_{p}^*}\zeta_p^{-z_1c}\sum_{z_2 \in \mathbb{F}_{p}^*}\zeta_p^{-z_2\rho}\sum_{x \in \mathbb{F}_{p^m}}\zeta_p^{{\rm Tr}(z_1x^{p^k+1}+z_2ax)}\sum_{y \in \mathbb{F}_{p^m}}\zeta_p^{{\rm Tr}(z_1y^{p^\ell+1}+z_2by)}\Bigg)-\frac{1}{p^2}\sum_{z_1,z_2 \in \mathbb{F}_{p}}\zeta_p^{-\left(z_1c+z_2\rho\right)}\\
&=p^{2m-2}+{\frac{1}{p^2}}(\Omega_1+\Omega_2)-\frac{1}{p^2}\sum_{z_1,z_2 \in \mathbb{F}_{p}}\zeta_p^{-\left(z_1c+z_2\rho\right)},
\end{split}
\end{equation}
where
\begin{equation}\label{eq:omega1}
\Omega_1=\sum_{z_1 \in \mathbb{F}_{p}^*}\zeta_p^{-z_1c}\sum_{x \in \mathbb{F}_{p^m}}\eta_m^{r_1}(z_1)\zeta_p^{{\rm Tr}(x^{p^k+1})}\sum_{y \in \mathbb{F}_{p^m}}\eta_m^{r_2}(z_1)\zeta_p^{{\rm Tr}(y^{p^\ell+1})} =\sum_{z_1 \in \mathbb{F}_{p}^*}\zeta_p^{-z_1c} S_k(1,0)S_\ell(1,0)
\end{equation}
since $r_1+r_2$ is always even and
\begin{equation}\label{eq:omega2}
\begin{split}
\Omega_2&=\sum_{z_1\in \mathbb{F}_{p}^*}\zeta_p^{-z_1c}\sum_{z_2\in \mathbb{F}_{p}^*}\zeta_p^{-z_2\rho}\sum_{x \in \mathbb{F}_{p^m}}\zeta_p^{{\rm Tr}(z_1x^{p^k+1}+z_2ax)}\sum_{y \in \mathbb{F}_{p^m}}\zeta_p^{{\rm Tr}(z_1y^{p^\ell+1}+z_2by)}\\
&=\sum_{z_1\in \mathbb{F}_{p}^*}\zeta_p^{-z_1c}\sum_{z_2\in \mathbb{F}_{p}^*}\zeta_p^{-z_2\rho}S_k(z_1,z_2a)S_\ell(z_1,z_2b).
\end{split}
\end{equation}
Here, $S_k(z_1,z_2a)$ and $S_\ell(z_1,z_2b)$ are defined in (\ref{eq:sab}). In order to determine the weight and complete  weight of a codeword in $\mathcal{C}_D$, we need to
calculate the values of $\Omega_1$ and $\Omega_2$. In the following, we give our main results.

\begin{theorem}\label{theorem1}
Let $\mathcal{C}_D$ be a linear code defined in (\ref{defcode}) with the defining set $D$ given in (\ref{defset}). Assume that $2v_2(m)=v_2(u)+v_2(v)$, then the following statements hold.

\noindent {\rm (1)} If c=0, then $\mathcal{C}_D$ is a $[p^{2m-1}+(-1)^{\frac{(p-1)m}{2}}(p-1)p^{m-1}-1,2m]$ two-weight linear code with weight enumerator
\[ 1+(p^{2m-1}+(-1)^{\frac{(p-1)m}{2}}(p-1)p^{m-1}-1) x^{(p-1)p^{2m-2}}+ (p-1)(p^{2m-1}-(-1)^{\frac{(p-1)m}{2}}p^{m-1})x^{(p-1)(p^{2m-2}+(-1)^{\frac{(p-1)m}{2}}p^{m-1})}\]
and its complete weight enumerator is
\[\begin{split}
&w_0^{p^{2m-1}+(-1)^{\frac{(p-1)m}{2}}(p-1)p^{m-1}-1}+(p^{2m-1}+(-1)^{\frac{(p-1)m}{2}}(p-1)p^{m-1}-1)
w_0^{p^{2m-2}+(-1)^{\frac{(p-1)m}{2}}(p-1)p^{m-1}-1}\prod_{\rho\in \mathbb{F}_{p}^*}w_\rho^{p^{2m-2}}\\
&+(p-1)(p^{2m-1}-(-1)^{\frac{(p-1)m}{2}}p^{m-1})w_0^{p^{2m-2}-1}\prod_{\rho\in \mathbb{F}_{p}^*}w_\rho^{p^{2m-2}+(-1)^{\frac{(p-1)m}{2}}p^{m-1}}.
\end{split}\]

\noindent {\rm (2)} Let $c\in \mathbb{F}_p^*$ and $g$ be a generator of $\mathbb{F}_p^*$, then $\mathcal{C}_D$ is a $[p^{2m-1}-(-1)^{\frac{(p-1)m}{2}}p^{m-1},2m]$ two-weight linear code with weight enumerator
$$1+(\frac{p+1}{2}p^{2m-1}+\frac{p-1}{2}(-1)^{\frac{(p-1)m}{2}}p^{m-1}-1)x^{(p-1)p^{2m-2}}
+\frac{p-1}{2}(p^{2m-1}-(-1)^{\frac{(p-1)m}{2}}p^{m-1})x^{(p-1)p^{2m-2}-(-1)^{\frac{(p-1)m}{2}}2p^{m-1}}$$
and its complete weight enumerator is
 \[ \begin{split}
&w_0^{p^{2m-1}-(-1)^{\frac{(p-1)m}{2}}p^{m-1}}+(p^{2m-1}+(-1)^{\frac{(p-1)m}{2}}(p-1)p^{m-1}-1)
w_0^{p^{2m-2}-(-1)^{\frac{(p-1)m}{2}}p^{m-1}}\prod_{\rho\in \mathbb{F}_{p}^*}w_\rho^{p^{2m-2}}+\\ &(p^{2m-1}-(-1)^{\frac{(p-1)m}{2}}p^{m-1})\Bigg(\sum_{j=1}^\frac{(p-1)}{2}w_0^{p^{2m-2}+(-1)^{\frac{(p-1)(m+1)}{2}}p^{m-1}}w_{2g^j}^{p^{2m-2}}w_{p-2g^j}^{p^{2m-2}}
\prod_{\mbox{\tiny$\begin{array}{c}
\rho\in \mathbb{F}_{p}^*\\
\rho\ne\pm2g^j
\end{array}$}}w_\rho^{p^{2m-2}+(-1)^{\frac{(p-1)m}{2}}p^{m-1}\eta(\rho^2-4g^{2j})}\\
&+\sum_{j=1}^\frac{(p-1)}{2}w_0^{p^{2m-2}-(-1)^{\frac{(p-1)(m+1)}{2}}p^{m-1}}
\prod_{\rho\in \mathbb{F}_{p}^*}w_\rho^{p^{2m-2}+(-1)^{\frac{(p-1)m}{2}}p^{m-1}\eta(\rho^2-4g^{2j+1})}\Bigg).
\end{split}\]
\end{theorem}

{\it Proof.}
We first determine the possible values of $\Omega_1$ and $\Omega_2$ which are defined in (\ref{eq:omega1}) and (\ref{eq:omega2}), respectively. By Proposition~\ref{pro:ww1}, in the case of $2v_2(m)=v_2(u)+v_2(v)$, the length of the linear code $\mathcal{C}_D$ is
\begin{equation*}
\begin{split}
n=\begin{cases}
p^{2m-1}-(-1)^{\frac{(p-1)m}{2}}p^{m-1}, & \text{if \,\,$c\in \mathbb{F}_p^*$,} \vspace{3mm} \\
p^{2m-1}+(-1)^{\frac{(p-1)m}{2}}(p-1)p^{m-1}-1, & \text{if \,\,$c=0$.} \\
\end{cases}
\end{split}
\end{equation*}

Recall that $u=\gcd(m,k)$ and $v=\gcd(m,\ell)$. Since $2v_2(m)=v_2(u)+v_2(v)$, we know that $v_2(m)=v_2(u)=v_2(v)$.
By Lemmas~\ref{lem:rank} and~\ref{lem:ska0} we have
 \begin{equation}
\begin{split}\label{eq:q1}
\Omega_1 =\sum_{z_1 \in \mathbb{F}_{p}^*}\zeta_p^{-z_1c}S_k(1,0) S_\ell(1,0) = \left\{ \begin{array}{lll}
-(-1)^{\frac{(p-1)m}{2}}p^{m}, &\quad {\rm if }\,\, c\in \mathbb{F}_p^*, \vspace{3mm} \\
(-1)^{\frac{(p-1)m}{2}}(p-1)p^{m}, &\quad {\rm if } \,\, c=0.\end{array} \right.
\end{split}
\end{equation}

In the following, we determine the possible values of $\Omega_2$. As $\frac{m}{u}$ and $\frac{m}{v}$ are odd, we verify that $f_k(x) =z_1^{p^{k}}x^{p^{2k}}+z_1 x$ and $f_\ell(x) = z_1^{p^{\ell}}x^{p^{2\ell}}+z_1 x$ are permutations over $\bF_{p^m}$ for any $z_1\in \bF_p^*$. Assume that $\gamma_a$ and $\gamma_b$ are the solutions of the equations $x^{p^{2k}}+x= -a^{p^k}$ and $x^{p^{2\ell}}+x= -b^{p^\ell}$, respectively. Then $z_1^{-1}\gamma_a z_2$ and $z_1^{-1}\gamma_bz_2$ are
the solutions of $f_k(x) = -(z_2a)^{p^k}$ and $f_\ell(x) = -(z_2b)^{p^\ell}$ for any $z_2\in \bF_p^*$, respectively.  By Lemma~\ref{lem:skab} we have
\begin{equation*}
\begin{split}
\Omega_2&=\sum_{z_1\in \mathbb{F}_{p}^*}\zeta_p^{-z_1c}\sum_{z_2\in \mathbb{F}_{p}^*}\zeta_p^{-z_2\rho}S_k(z_1,z_2a)S_\ell(z_1,z_2b)\\
&=(-1)^{\frac{(p-1)m}{2}}p^{m}\sum_{z_1,z_2\in \mathbb{F}_{p}^*}\zeta_p^{-z_1c-z_2\rho}\eta_m(z_1)\overline{\chi(z_1(z_1^{-1}\gamma_a z_2)^{p^k+1})}
\eta_m(z_1)\overline{\chi(z_1(z_1^{-1}\gamma_b z_2)^{p^\ell+1})}\\
&=(-1)^{\frac{(p-1)m}{2}}p^{m}\sum_{z_1\in \mathbb{F}_{p}^*}\zeta_p^{-z_1c}\sum_{z_2\in \mathbb{F}_{p}^*}\zeta_p^{-z_2\rho}\chi\left(-z_1^{-1}z_2^{2}\left(\gamma_a^{p^k+1}+\gamma_b^{p^\ell+1}\right)\right)\\
&=(-1)^{\frac{(p-1)m}{2}}p^{m}\sum_{z_1\in \mathbb{F}_{p}^*}\zeta_p^{-z_1c}\sum_{z_2\in \mathbb{F}_{p}^*}\zeta_p^{-\frac{z_2^{2}}{z_1}{{\rm Tr}\left(\gamma_a^{p^k+1}+\gamma_b^{p^\ell+1}\right)}-z_2\rho},
\end{split}
\end{equation*}
where $i=\sqrt{-1}$. If ${{\rm Tr}\left(\gamma_a^{p^k+1}+\gamma_b^{p^\ell+1}\right)}=0$, then
\begin{equation}\label{eq:a1}
\Omega_2(c,\rho)=(-1)^{\frac{(p-1)m}{2}}p^{m}\sum_{z_1\in \mathbb{F}_{p}^*}\zeta_p^{-z_1c}\sum_{z_2\in \mathbb{F}_{p}^*}\zeta_p^{-z_2\rho}.
\end{equation}
If ${{\rm Tr}_1^m(\gamma_a^{p^k+1}+\gamma_b^{p^\ell+1})}\ne0$, $c=0$ and $\rho=0$, then
\begin{equation}\label{eq:a3}
\Omega_2(0,0)=(-1)^{\frac{(p-1)m}{2}}p^{m}\sum_{z_1\in \mathbb{F}_{p}^*}\sum_{z_2\in \mathbb{F}_{p}^*}\zeta_p^{-\frac{z_2^{2}}{z_1}{{\rm Tr}(\gamma_a^{p^k+1}+\gamma_b^{p^\ell+1})}}=-(-1)^{\frac{(p-1)m}{2}}(p-1)p^{m}.
\end{equation}
If ${{\rm Tr}_1^m(\gamma_a^{p^k+1}+\gamma_b^{p^\ell+1})}\ne0$, $c=0$ and $\rho\in\mathbb{F}_{p}^*$, then
\begin{equation}\label{eq:a2}
\Omega_2(0,\rho)=(-1)^{\frac{(p-1)m}{2}}p^{m}\sum_{z_1\in \mathbb{F}_{p}^*}\sum_{z_2\in \mathbb{F}_{p}^*}\zeta_p^{-\frac{z_2^{2}}{z_1}{{\rm Tr}(\gamma_a^{p^k+1}+\gamma_b^{p^\ell+1})}-z_2\rho}
=(-1)^{\frac{(p-1)m}{2}}p^{m}.
\end{equation}
If ${{\rm Tr}_1^m(\gamma_a^{p^k+1}+\gamma_b^{p^\ell+1})}\ne0$, $c\in \mathbb{F}_p^*$ and $\rho=0$, from Lemma~\ref{lem:functions} we have that
\begin{equation}\label{eq:a4}
\begin{split}
\Omega_2(c,0)&=(-1)^{\frac{(p-1)m}{2}}p^{m}\sum_{z_1\in \mathbb{F}_{p}^*}\zeta_p^{-z_1c}\sum_{z_2\in \mathbb{F}_{p}^*}\zeta_p^{-\frac{z_2^{2}}{z_1}{{\rm Tr}(\gamma_a^{p^k+1}+\gamma_b^{p^\ell+1})}}\\
&=(-1)^{\frac{(p-1)m}{2}}p^{m}\sum_{z_1\in \mathbb{F}_{p}^*}\zeta_p^{-z_1c}\eta\left(-z_1^{-1}{\rm Tr}(\gamma_a^{p^k+1}+\gamma_b^{p^\ell+1})\right)G_1+(-1)^{\frac{(p-1)m}{2}}p^{m}\\
&=(-1)^{\frac{(p-1)m}{2}}p^{m}\eta\left(c{\rm Tr}(\gamma_a^{p^k+1}+\gamma_b^{p^\ell+1})\right){G_1^2}+(-1)^{\frac{(p-1)m}{2}}p^{m}\\
&=(-1)^{\frac{(p-1)m}{2}}p^{m+1}\eta\left(-c{\rm Tr}(\gamma_a^{p^k+1}+\gamma_b^{p^\ell+1})\right)+(-1)^{\frac{(p-1)m}{2}}p^{m}.
\end{split}
\end{equation}
If ${{\rm Tr}_1^m(\gamma_a^{p^k+1}+\gamma_b^{p^\ell+1})}\ne0$, $c\in \mathbb{F}_p^*$ and $\rho\in \mathbb{F}_p^*$, from Lemma~\ref{lem:functions} we have that
\begin{equation}\label{eq:a5}
\begin{split}
\Omega_2(c,\rho)&=(-1)^{\frac{(p-1)m}{2}}p^{m}\sum_{z_1\in \mathbb{F}_{p}^*}\zeta_p^{-z_1c}\sum_{z_2\in \mathbb{F}_{p}^*}\zeta_p^{-\frac{z_2^{2}}{z_1}{{\rm Tr}(\gamma_a^{p^k+1}+\gamma_b^{p^\ell+1})}-z_2\rho}\\
&=(-1)^{\frac{(p-1)m}{2}}p^{m}\sum_{z_1\in \mathbb{F}_{p}^*}\zeta_p^{-z_1c}\left(\sum_{z_2\in \mathbb{F}_{p}}\zeta_p^{-\frac{z_2^{2}}{z_1}{{\rm Tr}(\gamma_a^{p^k+1}+\gamma_b^{p^\ell+1})}-z_2\rho}-1\right)\\
&=(-1)^{\frac{(p-1)m}{2}}p^{m} \left( \eta\left(-{\rm Tr}(\gamma_a^{p^k+1}+\gamma_b^{p^\ell+1})\right){G_1}\sum_{z_1\in \mathbb{F}_{p}^*}\zeta_p^{\left(\frac{\rho^2}{{4\rm Tr}(\gamma_a^{p^k+1}+\gamma_b^{p^\ell+1})}-c\right)z_1}\eta(z_1)+1 \right)\\
&= \left\{ \begin{array}{lll}
(-1)^{\frac{(p-1)m}{2}}p^{m}, &\quad {\rm if }\,\,{\rm Tr}(\gamma_a^{p^k+1}+\gamma_b^{p^\ell+1})=\rho^2/4c , \vspace{3mm} \\
(-1)^{\frac{(p-1)m}{2}}p^{m}\left( p\eta\left(\rho^2-4c{\rm Tr}(\gamma_a^{p^k+1}+\gamma_b^{p^\ell+1})\right)+1\right), &\quad {\rm if } \,\, {\rm Tr}(\gamma_a^{p^k+1}+\gamma_b^{p^\ell+1})\ne \rho^2/4c.\end{array} \right.
\end{split}
\end{equation}
Let $\rho\in \bF_p^*$. From (\ref{eq:nab}), (\ref{eq:q1}) and (\ref{eq:a1})-(\ref{eq:a5}), we obtain the following results.

\noindent (I) If $c=0$, then
\begin{equation}\label{eq:casec=0}
\begin{split}
&N(0,0)=\left\{ \begin{array}{lc}
p^{2m-2}+(-1)^{\frac{(p-1)m}{2}}(p-1) p^{m-1}-1, & \text{if \,\,${{\rm Tr}(\gamma_a^{p^k+1}+\gamma_b^{p^\ell+1})}=0$,} \vspace{3mm} \\
p^{2m-2}-1, & \text{if \,\,${{\rm Tr}(\gamma_a^{p^k+1}+\gamma_b^{p^\ell+1})}\ne0$,}
\end{array}\right. \\ \\
&N(0,\rho)=\left\{ \begin{array}{lc}
p^{2m-2}, & \text{if \,\,${{\rm Tr}(\gamma_a^{p^k+1}+\gamma_b^{p^\ell+1})}=0$,} \vspace{3mm} \\
p^{2m-2}+(-1)^{\frac{(p-1)m}{2}}p^{m-1}, & \text{if \,\,${{\rm Tr}(\gamma_a^{p^k+1}+\gamma_b^{p^\ell+1})}\ne0${\red.}}
\end{array}\right.
\end{split}
\end{equation}

\noindent (II) If $c\in \mathbb{F}_p^*$, then
\begin{equation}\label{eq:casecneq0}
\begin{split}
&N(c,0)=\left\{ \begin{array}{lc}
p^{2m-2}-(-1)^{\frac{(p-1)m}{2}}p^{m-1}, & \text{if \,\,${{\rm Tr}(\gamma_a^{p^k+1}+\gamma_b^{p^\ell+1})}=0$,} \vspace{3mm} \\
p^{2m-2}+(-1)^{\frac{(p-1)m}{2}}p^{m-1}\eta\left(-c{\rm Tr}(\gamma_a^{p^k+1}+\gamma_b^{p^\ell+1})\right), & \text{if \,\,${{\rm Tr}(\gamma_a^{p^k+1}+\gamma_b^{p^\ell+1})}\ne0$,}
\end{array}\right. \\ \\
&N(c,\rho)= \left\{\begin{array}{ll}
p^{2m-2}, \hspace{-8mm} & \text{if \,${{\rm Tr}(\gamma_a^{p^k+1}+\gamma_b^{p^\ell+1})}=0$ or $\rho^2/4c$}, \vspace{3mm}\\
p^{2m-2}+(-1)^{\frac{(p-1)m}{2}}p^{m-1}\eta(\rho^2-4c{\rm Tr}(\gamma_a^{p^k+1}+\gamma_b^{p^\ell+1})), \hspace{-2mm} & \text{if \,${{\rm Tr}(\gamma_a^{p^k+1}+\gamma_b^{p^\ell+1})}\ne 0$ and $\rho^2/4c$.}
\end{array}\right.
\end{split}
\end{equation}

In the case of $c=0$, from (\ref{eq:wt}) and (\ref{eq:casec=0}), we know that for any $(a, b)\neq (0,0)$, the possible weight of $\mathbf{c}(a,b)$ is $w_1=(p-1)p^{2m-2}$ or $w_2 = (p-1)(p^{2m-2}+(-1)^{\frac{(p-1)m}{2}}p^{m-1})$.
This means that the dimension of $\mathcal{C}_D$ is $2m$.  It is easy to verify that the minimum distance of the dual
of $\C_D$ is greater than or equal to $2$ from the non-degenerate property of the trace function if $(0,0) \not\in D$. Let $A_i$ denote the number of codewords of weight $i$ in $\mathcal{C}$. From the first two Pless power moments identities we have
\begin{equation*}
\left\{\begin{array}{lcl}
A_{w_1}=(p^{2m-1}+(-1)^{\frac{(p-1)m}{2}}(p-1)p^{m-1}-1),\vspace{3mm}\\
A_{w_2}=(p-1)(p^{2m-1}-(-1)^{\frac{(p-1)m}{2}}p^{m-1}).
\end{array}\right.
\end{equation*}
So, from (\ref{eq:casec=0}) we obtain the weight and complete weight enumerators of $\C_D$.

In the case of $c\in \bF_p^*$,  from (\ref{eq:wt}) and (\ref{eq:casecneq0}) we know that for any $(a, b)\neq (0,0)$, the possible weight of $\mathbf{c}(a,b)$ is $w_1=(p-1)p^{2m-2}$ or $w_2 = (p-1)p^{2m-2}-(-1)^{\frac{(p-1)m}{2}}2p^{m-1}$. Analysis similar to that in the case of $c=0$ shows that
\begin{equation}\label{rrrewrews0}
\left\{\begin{array}{lcl}
A_{w_1}=(\frac{p+1}{2}p^{2m-1}+\frac{p-1}{2}(-1)^{\frac{(p-1)m}{2}}p^{m-1}-1),\vspace{3mm}\\
A_{w_2}=\frac{p-1}{2}(p^{2m-1}-(-1)^{\frac{(p-1)m}{2}}p^{m-1}).
\end{array}\right.
\end{equation}
From (\ref{eq:wt}) and (\ref{eq:casecneq0}) we have
\begin{equation}\label{rrrewrews}
A_{w_1}=\left|\left\{ (x,y)\in \mathbb{F}_{p^m}^2\setminus\{(0,0)\}\,\,:\,\,{{\rm Tr}(\gamma_a^{p^k+1}+\gamma_b^{p^\ell+1})}=0\,\, \text{ or}\,\, -c{{\rm Tr}(\gamma_a^{p^k+1}+\gamma_b^{p^\ell+1})}\in NSQ\right\}\right|
\end{equation}
 and
\begin{equation}\label{rrrewrews1}
A_{w_2}=\left|\left\{ (x,y)\in \mathbb{F}_{p^m}^2\setminus\{(0,0)\}\,\,:\,\,-c{{\rm Tr}(\gamma_a^{p^k+1}+\gamma_b^{p^\ell+1})}\in SQ\right\}\right|,
\end{equation}
where SQ and NSQ denote the sets of all square and non-square elements in $\mathbb{F}_p^*$, respectively.

 It is known that $f_k(x) =z_1^{p^{k}}x^{p^{2k}}+z_1 x$ and $f_\ell(x) = z_1^{p^{\ell}}x^{p^{2\ell}}+z_1 x$ are permutations over $\mathbb{F}_{p^m}$ for any $z_1\in\mathbb{F}_{p^m}^*$. Then the solutions $\gamma_a$ of $x^{p^{2k}}+x=-a^{p^{k}}$ and $\gamma_b$ of $x^{p^{2\ell}}+x=-b^{p^{\ell}}$  run through $\mathbb{F}_{p^m}$ as $a$ and $b$ run through $\mathbb{F}_{p^m}$, respectively. From calculations of the code length in (\ref{eq:6}), we know that the nonzero values of ${\rm Tr}(\gamma_a^{p^k+1}+\gamma_b^{p^\ell+1})$ are uniformly distributed in $\bF_p^*$ when $\gamma_a,\gamma_b$ run through $\mathbb{F}_{p^m}$. For any $t \in \mathbb{F}_p^*$, from (\ref{rrrewrews0})-(\ref{rrrewrews1}) we see
 $$\left|\left\{ (x,y)\in \mathbb{F}_{p^m}^2\setminus\{(0,0)\}\,\,:\,\, {\rm Tr}(\gamma_a^{p^k+1}+\gamma_b^{p^\ell+1})=t\right\}\right|=p^{2m-1}-(-1)^{\frac{(p-1)m}{2}}p^{m-1}$$
 and
 $$\left|\left\{ (x,y)\in \mathbb{F}_{p^m}^2\setminus\{(0,0)\}\,\,:\,\, {\rm Tr}(\gamma_a^{p^k+1}+\gamma_b^{p^\ell+1})=0\right\}\right|= p^{2m-1}+(-1)^{\frac{(p-1)m}{2}}(p-1)p^{m-1}-1.$$

Let $g$ be a generator of $\mathbb{F}_p^*$. Assume that $c{\rm Tr}(\gamma_a^{p^k+1}+\gamma_b^{p^\ell+1})$ is a square element in $\bF_p^*$ with the form $g^{2j}$ for some $j$. If ${\rm Tr}(\gamma_a^{p^k+1}+\gamma_b^{p^\ell+1}) = \rho^2/(4c)$, then $\rho=\pm 2g^j$. So, from (\ref{eq:casecneq0}) we obtain the weight and complete weight enumerators of $\C_D$.  $\square$

\begin{example}
Let $m=2$, $p=3$, $k=0$, $\ell=0$.
\begin{description}
\item{(1)} If $c\in \mathbb{F}_p^*$, then $\mathcal{C}_D$ has parameters $[24,4,12]$ with weight enumerator $1+24x^{12}+56x^{18}$ and complete weight enumerator
    $w_0^{24}+24w_0^{12}w_1^6w_2^6+56w_0^6w_1^9w_2^9.$
\item{(2)} If $c=0$. then $\mathcal{C}_D$ has parameters $[32,4,18]$ with weight enumerator $1+32x^{18}+48x^{24}$ and complete weight enumerator
    $w_0^{32}+32w_0^{14}w_1^9w_2^9+48w_0^8w_1^{12}w_2^{12}.$
\end{description}
Let $m=3$, $p=3$, $k=1$, $\ell=2$.
\begin{description}
\item{(3)} If $c\in \mathbb{F}_p^*$, then $\mathcal{C}_D$ has parameters $[252,6,162]$ with weight enumerator $1+476x^{162}+252x^{180}$  and complete weight enumerator
    $w_0^{252}+476w_0^{90}w_1^{81}w_2^{81}+252w_0^{72}w_1^{90}w_2^{90}.$
\item{(4)} If  $c=0$, then $\mathcal{C}_D$ has parameters $[224,6,144]$ with weight enumerator $1+504x^{144}+224x^{162}$  and complete weight enumerator
    $w_0^{224}+504w_0^{80}w_1^{72}w_2^{72}+224w_0^{62}w_1^{81}w_2^{81}.$
\end{description}
These results have been verified by Magma programs.
\end{example}

\begin{theorem}\label{theorem2}
Let $\mathcal{C}_D$ be a linear code defined in (\ref{defcode}) with the defining set $D$ given in (\ref{defset}). Assume that $2v_2(m)=v_2(u)+v_2(v)+1$, then the following statements hold.

\noindent {\rm (1)} If c=0, then $\mathcal{C}_D$ is a $[p^{2m-1}+(-1)^{\frac{(p-1)m}{4}}(p-1)p^{m-1}-1,2m]$ two-weight linear code with weight enumerator
\[ 1+(p^{2m-1}+(-1)^{\frac{(p-1)m}{4}}(p-1)p^{m-1}-1)x^{(p-1)p^{2m-2}}+(p-1)(p^{2m-1}-(-1)^{\frac{(p-1)m}{4}}p^{m-1})x^{(p-1)(p^{2m-2}+(-1)^{\frac{(p-1)m}{4}}p^{m-1})}\]
and its complete weight enumerator is
\[\begin{split}
&w_0^{p^{2m-1}+(-1)^{\frac{(p-1)m}{4}}(p-1)p^{m-1}-1}+(p^{2m-1}+(-1)^{\frac{(p-1)m}{4}}(p-1)p^{m-1}-1)
w_0^{p^{2m-2}+(-1)^{\frac{(p-1)m}{4}}(p-1)p^{m-1}-1}\prod_{\rho\in \mathbb{F}_{p}^*}w_\rho^{p^{2m-2}}\\
&+(p-1)(p^{2m-1}-(-1)^{\frac{(p-1)m}{4}}p^{m-1})w_0^{p^{2m-2}-1}\prod_{\rho\in \mathbb{F}_{p}^*}w_\rho^{p^{2m-2}+(-1)^{\frac{(p-1)m}{4}}p^{m-1}}.
\end{split}\]

\noindent {\rm (2)} Let $c\in \mathbb{F}_p^*$ and $g$ be a generator of $\mathbb{F}_p^*$, then $\mathcal{C}_D$ is a $[p^{2m-1}-(-1)^{\frac{(p-1)m}{4}}p^{m-1},2m]$ two-weight linear code with weight enumerator
$$1+(\frac{p+1}{2}p^{2m-1}+\frac{p-1}{2}(-1)^{\frac{(p-1)m}{4}}p^{m-1}-1)x^{(p-1)p^{2m-2}}
+\frac{p-1}{2}(p^{2m-1}-(-1)^{\frac{(p-1)m}{4}}p^{m-1})x^{(p-1)p^{2m-2}-(-1)^{\frac{(p-1)m}{4}}2p^{m-1}}$$
and its complete weight enumerator is
 \[ \begin{split}
&w_0^{p^{2m-1}-(-1)^{\frac{(p-1)m}{4}}p^{m-1}}+(p^{2m-1}+(-1)^{\frac{(p-1)m}{4}}(p-1)p^{m-1}-1)
w_0^{p^{2m-2}-(-1)^{\frac{(p-1)m}{4}}p^{m-1}}\prod_{\rho\in \mathbb{F}_{p}^*}w_\rho^{p^{2m-2}}+\\ &(p^{2m-1}-(-1)^{\frac{(p-1)m}{4}}p^{m-1})\Bigg(\sum_{j=1}^\frac{(p-1)}{2}w_0^{p^{2m-2}+(-1)^{\frac{(p-1)(m+2)}{4}}p^{m-1}}w_{2g^j}^{p^{2m-2}}w_{p-2g^j}^{p^{2m-2}}
\prod_{\mbox{\tiny$\begin{array}{c}
\rho\in \mathbb{F}_{p}^*\\
\rho\ne\pm2g^j
\end{array}$}}w_\rho^{p^{2m-2}+(-1)^{\frac{(p-1)m}{4}}p^{m-1}\eta(\rho^2-4g^{2j})}\\
&+\sum_{j=1}^\frac{(p-1)}{2}w_0^{p^{2m-2}-(-1)^{\frac{(p-1)(m+2)}{4}}p^{m-1}}
\prod_{\rho\in \mathbb{F}_{p}^*}w_\rho^{p^{2m-2}+(-1)^{\frac{(p-1)m}{4}}p^{m-1}\eta(\rho^2-4g^{2j+1})}\Bigg).
\end{split}\]
\end{theorem}

{\it Proof.}  Recall that $u=\gcd(m, k)$ and $v=\gcd(m,\ell)$. When $2v_2(m)=v_2(u)+v_2(v)+1$, then $m$ is even and there are two cases:
 \begin{equation*}
\begin{split}
\left\{ \begin{array}{lcl}
v_2(m)=v_2(u)\,\, \text{and}\,\, v_2(m)=v_2(v)+1,\vspace{3mm}\\
v_2(m)=v_2(v)\,\, \text{and}\,\,v_2(m)=v_2(u)+1.\end{array} \right.
\end{split}
\end{equation*}
 In the following, we only prove the case $v_2(m)=v_2(u)$ and $v_2(m)=v_2(v)+1$. The other cases can be shown similarly.
\vskip 6pt
Since $2v_2(m)=v_2(u)+v_2(v)+1$, Proposition \ref{pro:ww1} shows that the length of the code $\C_D$ is
\begin{equation*}
\begin{split}
n=\begin{cases}
p^{2m-1}-(-1)^{\frac{(p-1)m}{4}}p^{m-1}, & \text{if \,\,$c\in \mathbb{F}_p^*$,} \vspace{3mm} \\
p^{2m-1}+(-1)^{\frac{(p-1)m}{4}}(p-1)p^{m-1}-1, & \text{if \,\,$c=0$.} \\
\end{cases}
\end{split}
\end{equation*}

In order to determine the weight enumerator and complete weight enumerator of $\mathcal{C}_D$, as is shown in Theorem~\ref{theorem1}, we first compute the possible values of $\Omega_1$ and $\Omega_2$, which are given in (\ref{eq:omega1}) and (\ref{eq:omega2}), respectively.
When $v_2(m) = v_2(u)$ and $v_2(m) = v_2(v) + 1$, we have that $v_2(m)\leq v_2(k)$ and $v_2(m)= v_2(\ell)+1$. By Lemmas~\ref{lem:rank} and~\ref{lem:ska0} we have
\begin{equation}\label{eq:omega12}
\begin{split}
\Omega_1 =\sum_{z_1 \in \mathbb{F}_{p}^*}\zeta_p^{-z_1c}S_k(1,0) S_\ell(1,0) = \left\{ \begin{array}{lll}
-(-1)^{\frac{(p-1)m}{4}}p^{m}, &\quad {\rm if }\,\, c\in \mathbb{F}_p^*, \vspace{3mm} \\
(-1)^{\frac{(p-1)m}{4}}(p-1)p^{m}, &\quad {\rm if } \,\,  c=0 .\end{array} \right.
\end{split}
\end{equation}

Next, we determine the possible values of $\Omega_2$. Since $v_2(m) = v_2(u)$ and $v_2(m) = v_2(v) + 1$, it follows that $\frac{m}{u}$ and $\frac{m}{2v}$ are odd. By Lemma~\ref{lemGxn}, $f_k(x) =z_1^{p^{k}}x^{p^{2k}}+z_1 x$ and $f_\ell(x) = z_1^{p^{\ell}}x^{p^{2\ell}}+z_1 x$ are permutations over $\bF_{p^m}$ for any $z_1\in \bF_p^*$.
Let $\gamma_a$ and $\gamma_b$ be the solutions of the equations $x^{p^{2k}}+x= -a^{p^k}$ and $x^{p^{2\ell}}+x= -b^{p^\ell}$, respectively. Then $z_1^{-1}\gamma_a z_2$ and $z_1^{-1}\gamma_b z_2$ are
the solutions of $f_k(x) = -(z_2a)^{p^k}$ and $f_\ell(x) = -(z_2b)^{p^\ell}$ for any $z_2\in \bF_p^*$, respectively. By Lemma~\ref{lem:skab},
\begin{equation*}
\begin{split}
\Omega_2&=\sum_{z_1\in \mathbb{F}_{p}^*}\zeta_p^{-z_1c}\sum_{z_2\in \mathbb{F}_{p}^*}\zeta_p^{-z_2\rho}S_k(z_1,z_2a)S_\ell(z_1,z_2b)\\
&=\sum_{z_1,z_2\in \mathbb{F}_{p}^*}\zeta_p^{-z_1c-z_2\rho}\left( (-1)^{m-1}(\sqrt{-1})^{\frac{(p-1)^2 3m}{4}} p^{\frac{m}{2}}\eta_m(z_1)\overline{\chi(z_1(z_1^{-1}\gamma_a z_2)^{p^k+1})}\right)
\left((-1)^{\frac{m}{2v}}p^\frac{m}{2}\overline{\chi(z_1(z_1^{-1}\gamma_b z_2)^{p^\ell+1})}\right)\\
&=(-1)^{\frac{(p-1)m}{4}}p^{m}\sum_{z_1\in \mathbb{F}_{p}^*}\zeta_p^{-z_1c}\sum_{z_2\in \mathbb{F}_{p}^*}\zeta_p^{-\frac{z_2^{2}}{z_1}{{\rm Tr}\left(\gamma_a^{p^k+1}+\gamma_b^{p^\ell+1}\right)}-z_2\rho}.
\end{split}
\end{equation*}
The last equality follows from the facts that $m$ is even and $\frac{m}{2v}$ is odd. Let $\rho\in \bF_p^*$. By a similar analysis to (\ref{eq:a1})-(\ref{eq:a5}), we have the following results.

\noindent (I) If $c=0$, then
\begin{equation}\label{eq:q31}
\begin{split}
&\Omega_2(0,0)=\left\{ \begin{array}{lc}
(-1)^{\frac{(p-1)m}{4}}(p-1)^2p^{m}, & \text{if \,\,${{\rm Tr}(\gamma_a^{p^k+1}+\gamma_b^{p^\ell+1})}=0$,} \vspace{3mm} \\
-(-1)^{\frac{(p-1)m}{4}}(p-1)p^{m}, & \text{if \,\,${{\rm Tr}(\gamma_a^{p^k+1}+\gamma_b^{p^\ell+1})}\ne0$,}
\end{array}\right. \\ \\
&\Omega_2(0,\rho)=\left\{ \begin{array}{lc}
-(-1)^{\frac{(p-1)m}{4}}(p-1)p^{m}, & \text{if \,\,${{\rm Tr}(\gamma_a^{p^k+1}+\gamma_b^{p^\ell+1})}=0$,} \vspace{3mm} \\
(-1)^{\frac{(p-1)m}{4}}p^{m}, & \text{if \,\,${{\rm Tr}(\gamma_a^{p^k+1}+\gamma_b^{p^\ell+1})}\ne0$.}
\end{array}\right.
\end{split}
\end{equation}

\noindent (II) If $c\in \mathbb{F}_p^*$, then
\begin{equation}\label{eq:q32}
\begin{split}
&\Omega_2(c,0)=\left\{ \begin{array}{lc}
-(-1)^{\frac{(p-1)m}{4}}(p-1)p^{m}, & \text{if \,\,${{\rm Tr}(\gamma_a^{p^k+1}+\gamma_b^{p^\ell+1})}=0$,} \vspace{3mm} \\
(-1)^{\frac{(p-1)m}{4}}p^{m}\left( p\eta(-c{\rm Tr}(\gamma_a^{p^k+1}+\gamma_b^{p^\ell+1}))+1\right), & \text{if \,\,${{\rm Tr}(\gamma_a^{p^k+1}+\gamma_b^{p^\ell+1})}\ne0$,}
\end{array}\right. \\ \\
&\Omega_2(c,\rho)= \left\{\begin{array}{lll}
(-1)^{\frac{(p-1)m}{4}}p^{m}, & \text{if \,${{\rm Tr}(\gamma_a^{p^k+1}+\gamma_b^{p^\ell+1})}=0$ or $\rho^2/4c$}, \vspace{3mm}\\
(-1)^{\frac{(p-1)m}{4}}p^{m}\left(p \eta(\rho^2-4c{\rm Tr}(\gamma_a^{p^k+1}+\gamma_b^{p^\ell+1}))+1\right), & \text{if \,${{\rm Tr}(\gamma_a^{p^k+1}+\gamma_b^{p^\ell+1})}\ne 0$ and $\rho^2/4c$.}
\end{array}\right.
\end{split}
\end{equation}
From (\ref{eq:nab}) and (\ref{eq:omega12})-(\ref{eq:q32}), we have the following results.

\noindent (I) If $c=0$, then
\begin{equation}\label{eq:casec=0,3}
\begin{split}
&N(0,0)=\left\{ \begin{array}{lc}
p^{2m-2}+(-1)^{\frac{(p-1)m}{4}}(p-1) p^{m-1}-1, & \text{if \,\,${{\rm Tr}(\gamma_a^{p^k+1}+\gamma_b^{p^\ell+1})}=0$,} \vspace{3mm} \\
p^{2m-2}-1, & \text{if \,\,${{\rm Tr}(\gamma_a^{p^k+1}+\gamma_b^{p^\ell+1})}\ne0$,}
\end{array}\right. \\ \\
&N(0,\rho)=\left\{ \begin{array}{lc}
p^{2m-2}, & \text{if \,\,${{\rm Tr}(\gamma_a^{p^k+1}+\gamma_b^{p^\ell+1})}=0$,} \vspace{3mm} \\
p^{2m-2}+(-1)^{\frac{(p-1)m}{4}}p^{m-1}, & \text{if \,\,${{\rm Tr}(\gamma_a^{p^k+1}+\gamma_b^{p^\ell+1})}\ne0$.}
\end{array}\right.
\end{split}
\end{equation}

\noindent (II) If $c\in \mathbb{F}_p^*$, then
\begin{equation}\label{eq:casecneq0,4}
\begin{split}
&N(c,0)=\left\{ \begin{array}{lc}
p^{2m-2}-(-1)^{\frac{(p-1)m}{4}}p^{m-1}, & \text{if \,\,${{\rm Tr}(\gamma_a^{p^k+1}+\gamma_b^{p^\ell+1})}=0$,} \vspace{3mm} \\
p^{2m-2}+(-1)^{\frac{(p-1)m}{4}}p^{m-1}\eta\left(-c{\rm Tr}(\gamma_a^{p^k+1}+\gamma_b^{p^\ell+1})\right), & \text{if \,\,${{\rm Tr}(\gamma_a^{p^k+1}+\gamma_b^{p^\ell+1})}\ne0$,}
\end{array}\right. \\ \\
&N(c,\rho)= \left\{\begin{array}{lll}
p^{2m-2}, & \text{if \,${{\rm Tr}(\gamma_a^{p^k+1}+\gamma_b^{p^\ell+1})}=0$ or $\rho^2/4c$}, \vspace{3mm}\\
p^{2m-2}+(-1)^{\frac{(p-1)m}{4}}p^{m-1}\eta(\rho^2-4c{\rm Tr}(\gamma_a^{p^k+1}+\gamma_b^{p^\ell+1})), \hspace{-2mm} & \text{if \,${{\rm Tr}(\gamma_a^{p^k+1}+\gamma_b^{p^\ell+1})}\ne 0$ and $\rho^2/4c$.}
\end{array}\right.
\end{split}
\end{equation}

As is shown in Theorem~\ref{theorem1}, from (\ref{eq:casec=0,3}) and (\ref{eq:casecneq0,4}) we can obtain the weight and complete weight enumerators of $\C_D$. $\square$


\begin{remark}
When $v_2(m)=v_2(u)$ and $v_2(m)=v_2(v)+1$, the weight enumerator in (1) and (2) of Theorem~\ref{theorem2} is exact Table~1 and Table 3 in~\cite{ZhuXu2017}, respectively.
\end{remark}

\begin{example}
Let $m=4$, $p=3$, $k=0$, $\ell=2$.
\begin{description}
\item{(1)} If $c\in \mathbb{F}_p^*$, then $\mathcal{C}_D$ has parameters $[2160,8,1404]$ with weight enumerator $1+2160x^{1404}+4400x^{1458}$ and complete weight enumerator
    $w_0^{2160}+2160w_0^{756}w_1^{702}w_2^{702}+4400w_0^{702}w_1^{729}w_2^{729}.$
\item{(2)} If $c=0$, then $\mathcal{C}_D$ has parameters $[2240,8,1458]$ with weight enumerator $1+2240x^{1458}+4320x^{1512}$ and complete weight enumerator
    $w_0^{2240}+2240w_0^{782}w_1^{729}w_2^{729}+4320w_0^{728}w_1^{756}w_2^{756}.$
\end{description}
Let $m=2$, $p=3$, $k=1$, $\ell=0$.
\begin{description}
\item{(1)} If $c\in \mathbb{F}_p^*$, then $\mathcal{C}_D$ has parameters $[30,4,18]$ with weight enumerator $1+50x^{18}+30x^{24}$ and complete weight enumerator
    $w_0^{30}+50w_0^{12}w_1^{9}w_2^{9}+30w_0^{6}w_1^{12}w_2^{12}.$
\item{(2)} If $c=0$, then $\mathcal{C}_D$ has parameters $[20,4,12]$ with weight enumerator $1+60x^{12}+20x^{18}$ and complete weight enumerator
    $w_0^{20}+60w_0^{8}w_1^{6}w_2^{6}+20w_0^{2}w_1^{9}w_2^{9}.$
\end{description}
These results have been verified by Magma programs. The code $\C_D$ with parameters $[20,4,12]$ is optimal with respect to the tables of best codes known maintained at http://www.codetables.de.
\end{example}

\begin{theorem}\label{theorem3}
Let $\varepsilon_u$ and $\varepsilon_v$ be symbols defined in (\ref{equan03}) and $\mathcal{C}_D$ be a linear code defined in (\ref{defcode}) with the defining set $D$ given in (\ref{defset}). Assume that $2v_2(m)>v_2(u)+v_2(v)+1$, then the following statements hold.

\noindent {\rm (1)} If c=0, then $\mathcal{C}_D$ is a $[p^{2m-1}+p^{m+\varepsilon_uu+\varepsilon_vv}-p^{m+\varepsilon_uu+\varepsilon_vv-1}-1,2m]$ three-weight linear code with weight enumerator
\[\begin{split}
&1+(p^{2m}-p^{2m-2\varepsilon_uu-2\varepsilon_{v}v})x^{(p-1)(p^{2m-2}+p^{m+\varepsilon_uu+\varepsilon_{v}v-1}-p^{m+\varepsilon_uu+\varepsilon_{v}v-2})}
+(p^{m-\varepsilon_uu-\varepsilon_{v}v}-1)(p^{m-\varepsilon_uu-\varepsilon_{v}v-1}+1)x^{(p-1)p^{2m-2}}\\
&+(p-1)(p^{2m-2\varepsilon_uu-2\varepsilon_{v}v-1}-p^{m-\varepsilon_uu-\varepsilon_{v}v-1})x^{(p-1)(p^{2m-2}+p^{m+\varepsilon_uu+\varepsilon_{v}v-1})}
\end{split}\]
and its complete weight enumerator is
\[\begin{split}
&w_0^{p^{2m-1}+p^{m+\varepsilon_uu+\varepsilon_vv}-p^{m+\varepsilon_uu+\varepsilon_vv-1}-1}+(p^{2m}-p^{2m-2\varepsilon_uu-2\varepsilon_vv})
w_0^{p^{2m-2}+(p-1)p^{m+\varepsilon_uu+\varepsilon_vv-2}-1}\prod_{\rho\in \mathbb{F}_{p}^*}w_\rho^{p^{2m-2}+(p-1)p^{m+\varepsilon_uu+\varepsilon_vv-2}}\\
&+(p^{m-\varepsilon_uu-\varepsilon_{v}v}-1)(p^{m-\varepsilon_uu-\varepsilon_{v}v-1}+1)w_0^{p^{2m-2}+(p-1) p^{m+\varepsilon_uu+\varepsilon_vv-1}-1}\prod_{\rho\in \mathbb{F}_{p}^*}w_\rho^{p^{2m-2}}\\
&+(p-1)(p^{2m-2\varepsilon_uu-2\varepsilon_vv-1}-p^{m-\varepsilon_uu-\varepsilon_{v}v-1})w_0^{p^{2m-2}-1} \prod_{\rho\in \mathbb{F}_{p}^*}w_\rho^{p^{2m-2}+p^{m+\varepsilon_uu+\varepsilon_vv-1}}.
\end{split}\]

\noindent {\rm (2)} Let $c\in \mathbb{F}_p^*$ and $g$ be a generator of $\mathbb{F}_p^*$, then $\mathcal{C}_D$ is a $[p^{2m-1}-p^{m+\varepsilon_uu+\varepsilon_vv-1},2m]$ three-weight linear code with weight enumerator
\[\begin{split}
&1+(p^{2m}-p^{2m-2\varepsilon_uu-2\varepsilon_vv})x^{(p-1)(p^{2m-2}-p^{m+\varepsilon_{u}u+\varepsilon_{v}v-2})}
+(\frac{p+1}{2}p^{2m-2\varepsilon_uu-2\varepsilon_{v}v-1}+\frac{p-1}{2}p^{m-\varepsilon_uu-\varepsilon_{v}v-1}-1)x^{(p-1)p^{2m-2}}\\
&+\frac{p-1}{2}(p^{2m-2\varepsilon_uu-2\varepsilon_{v}v-1}-p^{m-\varepsilon_uu-\varepsilon_{v}v-1})x^{(p-1)p^{2m-2}-2p^{m+\varepsilon_uu+\varepsilon_{v}v-1}}
\end{split}\]
and its complete weight enumerator is
\[ \begin{split}
&w_0^{p^{2m-1}-p^{m+\varepsilon_uu+\varepsilon_vv-1}}+(p^{2m}-p^{2m-2\varepsilon_uu-2\varepsilon_vv})
w_0^{p^{2m-2}-p^{m+\varepsilon_uu+\varepsilon_vv-2}}\prod_{\rho\in \mathbb{F}_{p}^*}w_\rho^{p^{2m-2}-p^{m+\varepsilon_uu+\varepsilon_vv-2}}+\\ &(p^{2m-2\varepsilon_uu-2\varepsilon_vv-1}+(p-1)p^{m-\varepsilon_uu-\varepsilon_vv-1}-1)w_0^{p^{2m-2}-p^{m+\varepsilon_uu+\varepsilon_vv-1}}\prod_{\rho\in \mathbb{F}_{p}^*}w_\rho^{p^{2m-2}}+(p^{2m-2\varepsilon_uu-2\varepsilon_{v}v-1}-p^{m-\varepsilon_uu-\varepsilon_{v}v-1})\cdot\\
&\Bigg(\sum_{j=1}^\frac{(p-1)}{2}w_0^{p^{2m-2}+(-1)^\frac{p-1}{2}p^{m+\varepsilon_uu+\varepsilon_vv-1}}w_{2g^j}^{p^{2m-2}}w_{p-2g^j}^{p^{2m-2}}
\prod_{\mbox{\tiny$\begin{array}{c}
\rho\in \mathbb{F}_{p}^*\\
\rho\ne\pm2g^j
\end{array}$}}w_\rho^{p^{2m-2}+p^{m+\varepsilon_uu+\varepsilon_vv-1}\eta(\rho^2-4g^{2j})}\\
&+\sum_{j=1}^\frac{(p-1)}{2}w_0^{p^{2m-2}-(-1)^\frac{p-1}{2}p^{m+\varepsilon_uu+\varepsilon_vv-1}}\prod_{\rho\in \mathbb{F}_{p}^*}w_\rho^{p^{2m-2}+p^{m+\varepsilon_uu+\varepsilon_vv-1}\eta(\rho^2-4g^{2j+1})}\Bigg).
\end{split}\]
\end{theorem}

{\it Proof.} Recall that $u=\gcd(m,k)$ and $v=\gcd(m,\ell)$. When $2v_2(m)>v_2(u)+v_2(v)+1$, then $m$ is even and there are six cases:
 \begin{equation*}
\begin{split}
\left\{ \begin{array}{lcl}
v_2(m)=v_2(u)\,\, \text{and} \,\, v_2(m)>v_2(v)+1,\\
v_2(m)=v_2(v)\,\, \text{and} \,\,v_2(m)>v_2(u)+1,\\
v_2(m)=v_2(u)+1\,\, \text{and}\,\, v_2(m)=v_2(v)+1,\\
v_2(m)=v_2(u)+1\,\, \text{and} \,\,  v_2(m)>v_2(v)+1,\\
v_2(m)>v_2(u)+1\,\, \text{and} \,\, v_2(m)=v_2(v)+1,\\
v_2(m)>v_2(u)+1\,\, \text{and}\,\, v_2(m)>v_2(v)+1.\end{array} \right.
\end{split}
\end{equation*}
  In the following, we only prove the case $v_2(m)>v_2(u)+1$ and $v_2(m)>v_2(v)+1$. The other cases can be shown similarly.

\vskip 6pt
If $v_2(m)>v_2(u)+1$ and $v_2(m)>v_2(v)+1$, then $\varepsilon_u=1$ and $\varepsilon_v=1$ from~(\ref{equan03}). By Proposition \ref{pro:ww1},  the length of the code $\mathcal{C}_D$ is
\begin{equation*}
\begin{split}
n=\begin{cases}
p^{2m-1}-p^{m+u+v-1}, & \text{if $c\in \mathbb{F}_p^*$,} \vspace{3mm} \\
p^{2m-1}+p^{m+u+v}-p^{m+u+v-1}-1, & \text{if $c=0$.} \\
\end{cases}
\end{split}
\end{equation*}
We first determine the possible values of $\Omega_1$ and $\Omega_2$, which are given in (\ref{eq:omega1}) and (\ref{eq:omega2}), respectively.
\vskip 6pt
If $v_2(m) > v_2(u)+1$ and $v_2(m) > v_2(v) + 1$, then $v_2(m)> v_2(k)+1$ and $v_2(m)>v_2(\ell)+1$. By Lemmas~\ref{lem:rank} and~\ref{lem:ska0}, we have
\begin{equation}\label{eq:omega13}
\begin{split}
\Omega_1 =\sum_{z_1 \in \mathbb{F}_{p}^*}\zeta_p^{-z_1c}S_k(1,0) S_\ell(1,0) = \left\{ \begin{array}{lll}
-p^{m+u+v}, &\quad {\rm if }\,\, c\in \mathbb{F}_p^*, \vspace{3mm} \\
(p-1)p^{m+u+v}, &\quad {\rm if } \,\,  c=0 .\end{array} \right.
\end{split}
\end{equation}

In the following, we determine the possible values of $\Omega_2$. As $v_2(m) > v_2(u)+1$ and $v_2(m) > v_2(v) + 1$, by Lemma~\ref{lemGxn}, then
$f_k(x) =z_1^{p^{k}}x^{p^{2k}}+z_1 x$ and $f_\ell(x) = z_1^{p^{\ell}}x^{p^{2\ell}}+z_1 x$ are not permutations over $\bF_{p^m}$ for any
$z_1\in \bF_p^*$. If $f_k(x) =-(z_2a)^{p^{k}}$ has no solution in $\bF_{p^m}$ or $f_\ell(x) =-(z_2b)^{p^{\ell}}$ has no solution in
$\mathbb{F}_{p^m}$, then by Lemma~\ref{lem:skab2}, it is easy to see that
\begin{equation}\label{eq:omega231}
\Omega_2=0.
\end{equation}
Otherwise, assume that $\gamma_a$ and $\gamma_b$ are the solutions of the equations $x^{p^{2k}}+x= -a^{p^k}$ and $x^{p^{2\ell}}+x= -b^{p^\ell}$, respectively. Then $z_1^{-1}\gamma_a z_2$ and $z_1^{-1}\gamma_b z_2$ are
the solutions of $f_k(x) = -(z_2a)^{p^k}$ and $f_\ell(x) = -(z_2b)^{p^\ell}$ for any $z_2\in \bF_p^*$, respectively. By Lemma~\ref{lem:skab2},
we have
\begin{equation*}
\begin{split}
\Omega_2&=\sum_{z_1\in \mathbb{F}_{p}^*}\zeta_p^{-z_1c}\sum_{z_2\in \mathbb{F}_{p}^*}\zeta_p^{-z_2\rho}\left(-(-1)^{\frac{m}{2{u}}}p^{\frac{m}{2}+u}\overline{\chi(z_1(z_1^{-1}\gamma_a z_2)^{p^k+1})}\right)
\left(-(-1)^{\frac{m}{2{v}}}p^{\frac{m}{2}+v}\overline{\chi(z_1(z_1^{-1}\gamma_b z_2)^{p^\ell+1})}\right)\\
&=p^{m+u+v}\sum_{z_1\in \mathbb{F}_{p}^*}\zeta_p^{-z_1c}\sum_{z_2\in \mathbb{F}_{p}^*}\zeta_p^{-\frac{z_2^{2}}{z_1}{{\rm Tr}\left(\gamma_a^{p^k+1}+\gamma_b^{p^\ell+1}\right)}-z_2\rho}.
\end{split}
\end{equation*}
Let $\rho\in \bF_p^*$. By a similar analysis in (\ref{eq:a1})-(\ref{eq:a5}), we have the following results.

\noindent (I) If $c=0$, then
\begin{equation}\label{eq:q41}
\begin{split}
&\Omega_2(0,0)=\left\{ \begin{array}{lc}
(p-1)^2 p^{m+u+v}, & \text{if \,\,${{\rm Tr}(\gamma_a^{p^k+1}+\gamma_b^{p^\ell+1})}=0$,} \vspace{3mm} \\
-(p-1)p^{m+u+v}, & \text{if \,\,${{\rm Tr}(\gamma_a^{p^k+1}+\gamma_b^{p^\ell+1})}\ne0$,}
\end{array}\right. \\ \\
&\Omega_2(0,\rho)=\left\{ \begin{array}{lc}
-(p-1)p^{m+u+v}, & \text{if \,\,${{\rm Tr}(\gamma_a^{p^k+1}+\gamma_b^{p^\ell+1})}=0$,} \vspace{3mm} \\
p^{m+u+v}, & \text{if \,\,${{\rm Tr}(\gamma_a^{p^k+1}+\gamma_b^{p^\ell+1})}\ne0$.}
\end{array}\right.
\end{split}
\end{equation}

\noindent (II) If $c\in \mathbb{F}_p^*$, then
\begin{equation}\label{eq:q42}
\begin{split}
&\Omega_2(c,0)=\left\{ \begin{array}{lc}
-(p-1)p^{m+u+v}, & \text{if \,\,${{\rm Tr}(\gamma_a^{p^k+1}+\gamma_b^{p^\ell+1})}=0$,} \vspace{3mm} \\
p^{m+u+v}\left(p\eta(-c{\rm Tr}(\gamma_a^{p^k+1}+\gamma_b^{p^\ell+1}))+1\right), & \text{if \,\,${{\rm Tr}(\gamma_a^{p^k+1}+\gamma_b^{p^\ell+1})}\ne0$,}
\end{array}\right. \\ \\
&\Omega_2(c,\rho)= \left\{\begin{array}{lll}
p^{m+u+v}, & \text{if \,${{\rm Tr}(\gamma_a^{p^k+1}+\gamma_b^{p^\ell+1})}=0$ or $\rho^2/4c$}, \vspace{3mm}\\
p^{m+u+v}\left(p\eta(\rho^2-4c{\rm Tr}(\gamma_a^{p^k+1}+\gamma_b^{p^\ell+1}))+1\right), & \text{if \,${{\rm Tr}(\gamma_a^{p^k+1}+\gamma_b^{p^\ell+1})}\ne 0$ and $\rho^2/4c$.}
\end{array}\right.
\end{split}
\end{equation}
From (\ref{eq:nab}) and (\ref{eq:omega13})-(\ref{eq:q42}) we have the following results.

\noindent (I) If $c=0$ and one of $f_k(x) =-(z_2a)^{p^{k}}$ and $f_\ell(x) =-(z_2b)^{p^{\ell}}$ has no solution in $\bF_{p^m}$, then
\begin{equation}\label{eq:case01}
N(0,0)=p^{2m-2}+(p-1)p^{m+u+v-2}-1,\quad N(0,\rho)=p^{2m-2}+(p-1)p^{m+u+v-2}.
\end{equation}

\noindent (II) If $c=0$ and $f_k(x) =-(z_2a)^{p^{k}}$ and $f_\ell(x) =-(z_2b)^{p^{\ell}}$ has solutions in
$\mathbb{F}_{p^m}$, which are denoted by $z_1^{-1}\gamma_b z_2$ and $z_1^{-1}\gamma_a z_2$, respectively, then
\begin{equation}\label{eq:case02}
\begin{split}
&N(0,0)=\left\{ \begin{array}{lc}
p^{2m-2}+(p-1) p^{m+u+v-1}-1, & \text{if \,\,${{\rm Tr}(\gamma_a^{p^k+1}+\gamma_b^{p^\ell+1})}=0$,} \vspace{3mm} \\
p^{2m-2}-1, & \text{if \,\,${{\rm Tr}(\gamma_a^{p^k+1}+\gamma_b^{p^\ell+1})}\ne0$,}
\end{array}\right. \\ \\
&N(0,\rho)= \left\{\begin{array}{lc}
p^{2m-2}, \hspace{-8mm}& \text{if \,\,${{\rm Tr}(\gamma_a^{p^k+1}+\gamma_b^{p^\ell+1})}=0$,} \vspace{3mm}\\
p^{2m-2}+p^{m+u+v-1}, \hspace{-2mm} & \text{if \,\,${{\rm Tr}(\gamma_a^{p^k+1}+\gamma_b^{p^\ell+1})}\ne0$.}
\end{array}\right.
\end{split}
\end{equation}
\noindent (III) If $c\in \mathbb{F}_p^*$ and one of $f_k(x) =-(z_2a)^{p^{k}}$ and $f_\ell(x) =-(z_2b)^{p^{\ell}}$ has no solution in $\bF_{p^m}$, then
\begin{equation}\label{eq:case03}
N(c,0)=p^{2m-2}-p^{m+u+v-2},\quad N(c,\rho)=p^{2m-2}-p^{m+u+v-2}.
\end{equation}

\noindent (IV) If $c\in \mathbb{F}_p^*$ and $f_k(x) =-(z_2a)^{p^{k}}$ and $f_\ell(x) =-(z_2b)^{p^{\ell}}$
has solutions in $\mathbb{F}_{p^m}$, which are denoted by $z_1^{-1}\gamma_a z_2$ and $z_1^{-1}\gamma_b z_2$, respectively, then
\begin{equation}\label{eq:case04}
\begin{split}
&N(c,0)=\left\{ \begin{array}{lc}
p^{2m-2}-p^{m+u+v-1}, & \text{if \,\,${{\rm Tr}(\gamma_a^{p^k+1}+\gamma_b^{p^\ell+1})}=0$,} \vspace{3mm} \\
p^{2m-2}+p^{m+u+v-1}\eta\left(-c{\rm Tr}(\gamma_a^{p^k+1}+\gamma_b^{p^\ell+1})\right), & \text{if \,\,${{\rm Tr}(\gamma_a^{p^k+1}+\gamma_b^{p^\ell+1})}\ne0$,}
\end{array}\right. \\ \\
&N(c,\rho)= \left\{\begin{array}{lll}
p^{2m-2}, \hspace{-8mm}& \text{if \, ${{\rm Tr}(\gamma_a^{p^k+1}+\gamma_b^{p^\ell+1})}=0$ or $\rho^2/4c$}, \vspace{3mm}\\
p^{2m-2}+p^{m+u+v-1}\eta\left(\rho^2-4c{\rm Tr}(\gamma_a^{p^k+1}+\gamma_b^{p^\ell+1})\right), \hspace{-2mm} & \text{if \, ${{\rm Tr}(\gamma_a^{p^k+1}+\gamma_b^{p^\ell+1})}\ne0$ and $\rho^2/4c$}.
\end{array}\right.
\end{split}
\end{equation}

In the case of $c=0$, from (\ref{eq:wt}), (\ref{eq:case01}) and (\ref{eq:case02}) we known that the possible weight of $\mathbf{c}(a,b)$ is $w_1=(p-1)(p^{2m-2}+p^{m+u+v-1}-p^{m+u+v-2})$, $w_2=(p-1)p^{2m-2}$ or
$ w_3=(p-1)(p^{2m-2}+p^{m+u+v-1})$. This means that the dimension of $\mathcal{C}_D$ is $2m$. Let $A_i$ denote the number of codewords of weight $i$ in $\mathcal{C}_D$. To determine the multiplicity of each weight, we first investigate the multiplicity $A_{w_1}$ of the weight $w_1$. From above analysis, we know that
$\Omega_2=0$ if and only if $f_k(x) =-(z_2a)^{p^{k}}$ or $f_\ell(x) =-(z_2b)^{p^{\ell}}$ has no solution in $\bF_{p^m}$,  which is equivalent to that
$x^{p^{2k}} + x = -a^{p^k}$ or $x^{p^{2\ell}} + x = -b^{p^\ell}$ has no solution in $\bF_{p^m}$  since $z_1\in \bF_p^*$. So, by Lemma~\ref{lem:numsl} we have
\begin{equation*}
\begin{split}
&A_{w_1}=\left|\left\{(a,b)\in \mathbb{F}_{p^m}^2\setminus\{(0,0)\}:\,\,x^{p^{2k}} + x = -a^{p^k}\,\, and\,\,x^{p^{2\ell}} + x = -b^{p^\ell}\,\,{ \rm has\,\, no \,\,solution \,\,in}\,\,\mathbb{F}_{p^m}\right\}\right|\\
&+\left|\left\{(a,b)\in \mathbb{F}_{p^m}^2\setminus\{(0,0)\}:\,\,x^{p^{2k}} + x = -a^{p^k}\,\,{\rm has\,\,solutions \,\,in} \,\,\mathbb{F}_{p^m}\,\,{ \rm and}\,\,x^{p^{2\ell}} + x = -b^{p^\ell}\,\,{\rm has\,\,no \,\,solution \,\,in}\,\,\mathbb{F}_{p^m}\right\}\right|\\
&+\left|\left\{(a,b)\in \mathbb{F}_{p^m}^2\setminus\{(0,0)\}:\,\,x^{p^{2k}} + x = -a^{p^k}\,\,{\rm has\,\, no \,\,solution \,\,in} \,\,\mathbb{F}_{p^m}\,\,{\rm and}\,\,x^{p^{2\ell}} + x = -b^{p^\ell}\,\,{ \rm has\,\,solutions \,\, in} \,\,\mathbb{F}_{p^m}\right\}\right|\\
&=(p^m-p^{m-2u})(p^m-p^{m-2v})+(p^m-p^{m-2v})p^{m-2u}+(p^m-p^{m-2u})p^{m-2v}\\
&=p^{2m}-p^{2(m-u-v)}.
\end{split}
\end{equation*}
It is easy to verify that the minimum distance of the dual of $\C_D$ is greater than or equal to $2$  if $(0,0) \not\in D$. From the first two Pless power moments identities we have
\begin{equation*}
\left\{\begin{array}{lcl}
A_{w_1}=(p^{2m}-p^{2m-2u-2v}),\\
A_{w_2}=(p^{m-u-v}-1)(p^{m-u-v-1}+1),\\
A_{w_3}=(p-1)(p^{2m-2u-2v-1}-p^{m-u-v-1}).
\end{array}\right.
\end{equation*}
So, from (\ref{eq:case01}) and (\ref{eq:case02}) we obtain the weight and complete weight enumerators of $\C_D$.

In the case of $c\in \bF_p^*$, as is shown in Theorem~\ref{theorem1} and combining (\ref{eq:case03}) and (\ref{eq:case04}), we obtain the weight and complete weight enumerators of $\C_D$. $\square$

\begin{remark}
When $v_2(m)=v_2(u)$ and $v_2(m)>v_2(v)+1$, the weight enumerator in (1) and (2) of Theorem~\ref{theorem3} is exact Table~2 and Table~4 in~\cite{ZhuXu2017}, respectively.
\end{remark}

\begin{example}
Let $m=4$, $p=3$.
\begin{description}
\item{(1)} If $k=3$, $\ell=1$ and $c\in \mathbb{F}_p^*$, then $\varepsilon_u=\varepsilon_v=1$ and $\mathcal{C}_D$ has parameters $[1944,8,972]$ with weight enumerator $1+24x^{972}+6480x^{1296}+56x^{1458}$ and complete weight enumerator
    $w_0^{1944}+24w_0^{972}w_1^{486}w_2^{486}+6480w_0^{648}w_1^{648}w_2^{648}+56w_0^{486}w_1^{729}w_2^{729}.$
\item{(2)} If $k=1$, $\ell=2$ and $c=0$, then $\varepsilon_u=1$, $\varepsilon_v=0$ and $\mathcal{C}_D$ has parameters $[2348,8,1458]$ with weight enumerator $1+260x^{1458}+5832x^{1566}+468x^{1620}$ and complete weight enumerator
    $w_0^{2348}+260w_0^{890}w_1^{729}w_2^{729}+5832w_0^{782}w_1^{783}w_2^{783}+468w_0^{728}w_1^{810}w_2^{810}.$
\item{(3)} If $k=2$, $\ell=3$ and $c\in \mathbb{F}_p^*$, then $\varepsilon_u=0$, $\varepsilon_v=1$ and $\mathcal{C}_D$ has parameters $[2106,8,1296]$ with weight enumerator $1+234x^{1296}+5832x^{1404}+494x^{1458}$ and complete weight enumerator
    $w_0^{2160}+234w_0^{810}w_1^{648}w_2^{648}+5832w_0^{702}w_1^{702}w_2^{702}+494w_0^{648}w_1^{729}w_2^{729}.$
\item{(4)} If $k=2$, $\ell=2$ and $c=0$, then $\varepsilon_u=\varepsilon_v=0$ and $\mathcal{C}_D$ has parameters $[2240,8,1458]$ with weight enumerator $1+2240x^{1458}+4320x^{1512}$ and complete weight enumerator
    $w_0^{2240}+2240w_0^{782}w_1^{729}w_2^{729}+4320w_0^{728}w_1^{756}w_2^{756}.$
\end{description}
These results have been verified by Magma programs.
\end{example}

\section{Punctured codes of $\mathcal{C}_{D}$}

In this section, we investigate the punctured code $\mathcal{C}_{\bar{D}}$,  which is derived from $\mathcal{C}_{D}$ by deleting some coordinates of codewords in $\C_D$.
Some new two-weight and three-weight linear codes are obtained.

From Theorem \ref{theorem1}, Theorem \ref{theorem2} and Theorem \ref{theorem3}, it is observed that the Hamming weight of each codeword in $\mathcal{C}_{D}$ has a common divisor $p-1$ for $c=0$. This indicates that $\mathcal{C}_{D}$ may be punctured into a shorter one whose weight distribution is derived from that of the original code.
To this end, we define an equivalence relation in the set $D$ as follows. For $(\beta, \gamma), (\delta, \eta) \in D$, we say that $(\beta, \gamma)$ is equivalent to $(\delta, \eta)$ if and only if there exists $a\in \bF_p^*$ such that $(\delta, \eta) = a(\beta, \gamma)$. The elements chosen from each equivalent class in $D$ consist of a set $\bar{D}$. It is clear that
\begin{equation}\label{eq:barDf}
D=\mathbb{F}_p^*\bar{D}=\left\{z(x,y)=(zx,zy):z\in \mathbb{F}_p^*, (x,y)\in \bar{D}\right\}.
\end{equation}
Then the linear code $\mathcal{C}_{\bar{D}}$ defined in (\ref{defcode}) with the defining set $\bar{D}$ is a punctured version of $\mathcal{C}_{D}$, whose parameters are given in the following theorem.

\begin{theorem}\label{theorem4}
Let $\mathcal{C}_{\bar{D}}$ be the linear code defined as above, where $\bar{D}$ is defined in (\ref{eq:barDf}). Then the following statements hold.
\begin{description}
\item{{\rm (1)}} If $2v_2(m)=v_2(u)+v_2(v)$, then $\mathcal{C}_{\bar{D}}$ is a $\left[\frac{p^{2m-1}-1}{p-1}+(-1)^{\frac{(p-1)m}{2}}p^{m-1},2m\right]$ two-weight linear code with the weight distribution given in Table~\ref{Table7}.
\begin{table}[h]
{\caption{\rm   The weight distribution of $\mathcal{C}_{\bar{D}}$ for $2v_2(m)=v_2(u)+v_2(v)$ }\label{Table7}
\begin{center}
\begin{tabular}{cccc}\hline
     Weight & Multiplicity \\\hline
  $0$ & $1$  \\
 $p^{2m-2}$ & $p^{2m-1}+(-1)^{\frac{(p-1)m}{2}}(p-1)p^{m-1}-1$ \\
  $p^{2m-2}+(-1)^{\frac{(p-1)m}{2}}p^{m-1}$ & $(p-1)\left(p^{2m-1}-(-1)^{\frac{(p-1)m}{2}}p^{m-1}\right)$ \\
     \hline
\end{tabular}
\end{center}}
\end{table}
\item{{\rm (2)}} If $2v_2(m)=v_2(u)+v_2(v)+1$, then $\mathcal{C}_{\bar{D}}$ is a $\left[\frac{p^{2m-1}-1}{p-1}+(-1)^{\frac{(p-1)m}{4}}p^{m-1},2m\right]$ two-weight linear code with the weight distribution given in Table~\ref{Table8}.
\begin{table}
{\caption{\rm   The weight distribution of $\mathcal{C}_{\bar{D}}$ for $2v_2(m)=v_2(u)+v_2(v)+1$} \label{Table8}
\begin{center}
\begin{tabular}{cccc}\hline
     Weight & Multiplicity \\\hline
  $0$ & $1$  \\
 $p^{2m-2}$ & $p^{2m-1}+(-1)^{\frac{(p-1)m}{4}}(p-1)p^{m-1}-1$ \\
  $p^{2m-2}+(-1)^{\frac{(p-1)m}{4}}p^{m-1}$ & $(p-1)\left(p^{2m-1}-(-1)^{\frac{(p-1)m}{4}}p^{m-1}\right)$ \\
     \hline
\end{tabular}
\end{center}}
\end{table}
\item{{\rm (3)}} If $2v_2(m)>v_2(u)+v_2(v)+1$, then $\mathcal{C}_{\bar{D}}$ is a $\left[\frac{p^{2m-1}-1}{p-1}+p^{m+\varepsilon_uu+\varepsilon_vv-1},2m\right]$ three-weight linear code with the weight distribution given in Table~\ref{Table9}.
\begin{table}
{\caption{\rm   The weight distribution of $\mathcal{C}_{\bar{D}}$ for $2v_2(m)>v_2(u)+v_2(v)+1$}\label{Table9}
\begin{center}
\begin{tabular}{cccc}\hline
     Weight & Multiplicity \\\hline
  $0$ & $1$  \\
  $p^{2m-2}+p^{m+\varepsilon_uu+\varepsilon_{v}v-1}-p^{m+\varepsilon_uu+\varepsilon_{v}v-2}$ & $p^{2m}-p^{2m-2\varepsilon_uu-2\varepsilon_{v}v}$ \\
  $p^{2m-2}$ & $(p^{m-\varepsilon_uu-\varepsilon_{v}v}-1)(p^{m-\varepsilon_uu-\varepsilon_{v}v-1}+1)$ \\
  $p^{2m-2}+p^{m+\varepsilon_uu+\varepsilon_{v}v-1}$ & $(p-1)\left(p^{2m-2\varepsilon_uu-2\varepsilon_{v}v-1}-p^{m-\varepsilon_uu-\varepsilon_{v}v-1}\right)$ \\
     \hline
\end{tabular}
\end{center}}
\end{table}
\end{description}
\end{theorem}

\begin{remark}
Assume that $m=2$, $v_2(u)+v_2(v)=1$ and $p \equiv 3\pmod 4$. From Table \ref{Table8}, it is easy to see that $\mathcal{C}_{\bar{D}}$ is a $\left[p^2+1,4,p^2-p\right]$ code. From Chapter $13$ in \cite{Ding2019}, this code is called an ovoid code, which is optimal with respect to the Griesmer bound.
\end{remark}

\begin{example}
Let $m=2$ and $p=3$.
\begin{description}
\item{(1)} If $k=0$ and $\ell=0$, or $k=1$ and $\ell=1$, then $\mathcal{C}_{\bar{D}}$ has parameters $[16,4,9]$ and weight enumerator $1+32x^{9}+48x^{12}$.
\item{(2)} If $k=0$ and $\ell=1$, or $k=1$ and $\ell=0$, then $\mathcal{C}_{\bar{D}}$ has parameters $[10,4,6]$ and weight enumerator $1+60x^{6}+20x^{9}$.
\end{description}
These results have been verified by Magma programs.
All of these codes are optimal with respect to the tables of best codes known maintained at http://www.codetables.de.
\end{example}

\section{Conclusions }

This paper further studied a class of linear codes $\C_D$ proposed by Zhu and Xu~\cite{ZhuXu2017}, and determined their weight enumerators for all non-negative integers $m, k$ and $\ell$,
and generalized some results in~\cite{Jian2019,ZhuXu2017}. Moreover, we obtained the complete weight enumerators of $\C_D$ and the weight distribution of the punctured linear code
$\mathcal{C}_{\bar{D}}$ which is derived from $\mathcal{C}_{D}$ by deleting some coordinates. Some optimal and almost optimal linear codes are obtained .
These new two-weight and three-weight linear codes $\C_D$ may be applied to construct strongly regular graphs~\cite{Calderbank1986} and association schemes~\cite{Calderbank1984} with new parameters, respectively.
Furthermore, if $m$ is properly chosen, one easily check that
\[ \frac{w_{min}}{w_{max}} > \frac{p-1}{p}\]
for the linear codes $\C_D$, where $w_{min}$ and $w_{max}$ denote the minimum and maximum nonzero weights of $\C_D$, respectively. Then the new codes may be used to construct secret sharing
schemes with nice access structures~\cite{DDing2015}.


\begin {thebibliography}{100}



\bibitem{AhnKaLi2017} J.\ Ahn, D.\ Ka, C.\ Li, Complete weight enumerators of a class of linear codes, Des. Codes Cryptogr. 83 (2017) 83-99.

\bibitem{Blake1991}  I.F.\ Blake, K.\ Kith, On the complete weight enumerator of Reed-Solomon codes. SIAM J. Discrete Math. 4(2)(1991) 164-171.


\bibitem{Calderbank1984} A. R. Calderbank, J. M. Geothals, Three-weight codes and association schemes, Philips J. Res. 39 (1984) 143-152.

\bibitem{Calderbank1986} A. R. Calderbank, W. M. Kantor, The geometry of two-weight codes, Bull. Lond. Math. Soc. 18 (1986) 97-122.

\bibitem{Coulter19980} R. S. Coulter, Explicit evaluations of some Weil sums, Acta Arith. 83 (1998) 241-251.

\bibitem{Coulter1998} R. S. Coulter, Further evaluations of Weil sums, Acta Arith. 86 (1998) 217-226.

\bibitem{Ding2019} C. Ding, Designs from linear codes, World Scientific, Singapore (2019).

\bibitem{Ding2015} C. Ding, Linear codes from some 2-designs, IEEE Trans. Inf. Theory 61 (2015) 3265-3275.

\bibitem{Ding2016} C. Ding, A construction of binary linear codes from Boolean functions, Discrete Math. 339 (2016) 2288-2303.

\bibitem{DingNieder2007} C. Ding, H. Niederreiter, Cyclotomic linear codes of order 3, IEEE Trans. Inf. Theory 53 (2007) 2274-2277.

\bibitem{DingLietal2015} C. Ding, C. Li, N. Li  Z. Zhou, Three-weight cyclic codes and their weight distributions, Discrete Math. 339 (2016) 415-427.

\bibitem{DingC2008} C. Ding, Optimal constant composition codes from zero-dikerence balanced functions, IEEE Trans. Inf. Theory 54 (2008) 5766-7259.

\bibitem{DingWang2005} C. Ding, X. Wang, A coding theory construction of new systematic authentication codes, Theortical Computer Science 330 (2005) 81-99.

\bibitem{DingHellesethetal2007} C. Ding, T. Helleseth, T. Kl${\o}$ve, X. Wang,  A general construction of authentication codes, IEEE Trans. Inf. Theory  53 (2007) 2229-223.

\bibitem{DingYin2005} C. Ding, J. Yin, Algebraic constructions of constant composition codes. IEEE Trans. Inf. Theory 51 (2005) 1585-1589.


\bibitem{DDing2015} K. Ding, C. Ding, A class of two-weight and three-weight codes and their applications in secret sharing. IEEE Trans. Inf. Theory 61 (2015) 5835-5842.

\bibitem{Draper2007} S. Draper, X. Hou, Explicit evaluation of certain exponential sums of quadratic functions over $\mathbb{F}_{p^n}$, $p$ odd, arxiv:0708.3619v1. (2007).


\bibitem{HengYue2016} Z. Heng, Q. Yue, Two classes of two-weight linear codes, Finite Fields Appl. 38 (2016) 72-92.

\bibitem{HengYueLi2016} Z. Heng, Q. Yue, C. Li,  Three classes of linear codes with two or three weights, Discrete Math. 339 (2016) 2832-2847.

\bibitem{HengYue2017} Z. Heng, Q. Yue, A construction of $q$-ary linear codes with two weights, Finite Fields Appl. 48 (2017) 20-42.

\bibitem{HengYue20162} Z. Heng, Q. Yue, Evaluation of the Hamming weights of a classes of linear codes based on Gauss sums,  Des. Codes Cryptogr. 83 (2017) 307-326.


\bibitem{Jian2019} G. Jian, Z. Lin, R. Feng, Two-weight and three-weight linear codes based on Weil sums, Finite Fields Appl. 57 (2019) 92-107.

\bibitem{Klove2007} T. Kl{\o}ve, Codes for Error Detection, Hackensack, NJ: world Scientific, 2007.

\bibitem{Kuzmin2001} A.S.\ Kuzmin, A.A.\ Nechaev, Complete weight enumerators of generalized Kerdock code and related linear codes over Galois rings, Discrete Appl. Math. 111 (2001) 117-137.

\bibitem{LiYueFu2016} C.\ Li, Q.\ Yue, F.\ Fu, Complete weight enumerators of some cyclic codes,  Des. Codes Cryptogr. 80(2016) 295-315.


\bibitem{LiBae2016} C. Li, S. Bae, J. Ahn, S. Yang, Z. Yao, Complete weight enumerators of some linear codes and their applications, Des. Codes Cryptogr. 81 (2016) 153-168.

\bibitem{Lietal2016} F. Li, Q. Wang, D. Lin, A class of three-weight and five-weight linear codes, Discrete Appl. Math. 241 (2018) 25-38.

\bibitem{LuoCaoetal2018} G. Luo, X. Cao, S. Xu, J. Mi, Binary linear codes with two or three weights from niho exponents, Cryptogr. Commun. 10  (2018) 301-318.

\bibitem{LuoG2018} G. Luo, X. Cao, Complete weight enumerators of three classes of linear codes, Cryptogr. Commun. 10 (2018) 1091-1108.


\bibitem{Lidl1983} R. Lidl, H. Niederreiter, Finite Fields, Encyclopedia of Mathematics, Vol. 20, Cambridge University Press, Cambridge, 1983.

\bibitem{LiuYi2018} Y. Liu, Z. Liu, Complete weight enumerators of a new class linear codes, Discrete Math. 341 (2018) 1959-1972.

\bibitem{MacWilliam1997} F.J. MacWilliams, N.J.A. Sloane, The Theory of Error-Correcting Codes, North-Holland Publishing Company, 1997.

\bibitem{Tang2018} C. Tang, Y. Qi, D. Huang, Two-weight and three-weight linear codes from square functions,  IEEE Commun. Lett. 20 (2015) 29-32.

\bibitem{TangLietal2016} C. Tang, N. Li, Y. Qi, Z. Zhou, T. Helleseth, Linear codes with two or three weights from weakly regular bent functions, IEEE Trans. Inf. Theory 62 (2016) 1166-1176.

\bibitem{Tan2018} P. Tan, Z. Zhou, D. Tang, T. Helleseth,  The weight distribution of a class of two-weight linear codes derived from Kloosterman sums, Cryptogr. Commun. 10 (2018) 291-299.

\bibitem{HellesethKholosha2006} T.\ Helleseth, A.\ Kholosha, Monomial and quadratic bent functions over the finite fields of odd characteristic, IEEE Trans. Inf. Theory 52 (2006) 2018-2032.

\bibitem{WangQ2017} Q. Wang, F. Li, Complete weight enumerators of two classes of linear of linear codes, Discrete Math. 340 (2017) 467-480.

\bibitem{Wangetal2016} X. Wang, D. Zheng, H. Liu, Several classes of linear codes and their weight distributions, Appl. Algebra Eng. Commun. Comput. 30 (2019) 75-92.

\bibitem{WuY2019} Y. Wu, Q. Yue, X. Zhu, S. Yang, Weight enumerators of reducible cyclic codes and their dual codes, Discrete Math. 342 (2019) 671-682.

\bibitem{XuGuang2018} G. Xu, X. Cao, S. Xu, J. Ping, Complete weight enumerators of a class of linear codes with two weights, Discrete Math. 341 (2018) 525-535.

\bibitem{Xiaetal2017} Y. Xia, C. Li, Three-weight ternary linear codes from a family of power functions, Finite Fields Appl. 46 (2017) 17-37.

\bibitem{YangShu2019} S. Yang, Q. Yue, S. Yan, X. Kong, Complete weight enumerators of a class of two-weight linear codes, Cryptogr. Commun. 11 (2019) 609-620.


\bibitem{YangKong2017} S. Yang, X. Kong, C. Tang, A construction of linear codes and their complete weight enumerators, Finite Fields Appl. 48 (2017) 196-226.

\bibitem{ZhengBao2017} D. Zheng, J. Bao,  Four classes of linear codes from cyclotomic cosets, Des. Codes Cryptogr. 86 (2018) 1007-1022.

\bibitem{ZhouDing2013} Z. Zhou, C. Ding, Seven classes of three-weight cyclic codes, IEEE Trans. Inf. Theory 61 (2013) 4120-4126.

\bibitem{ZhouDing2014} Z. Zhou, C. Ding, A class of three-weight cyclic codes,  Finite Fields Appl. 25 (2014) 79-93.

\bibitem{ZhouLietal2015} Z. Zhou, N. Li, C. Fan, T. Helleseth, Linear codes with two or three weights from quadratic bent functions, Des. Codes Cryptogr. 81 (2015) 1-13.
\bibitem{ZhuXu2017} X. Zhu, D. Xu, Linear codes with two weights or three weights from two types of quadratic forms, Transactions of Nanjing University of Aeronaytics 34 (2017) 62-71.



\noindent
\end {thebibliography}
\end{document}